\documentclass[a4paper,12pt,twoside]{book}
\usepackage{enteteProjet}
\usepackage{bigints}
\newcommand{\matlabscript}[2]
  {\begin{itemize}\item[]\lstinputlisting[caption=#2,label=#1]{#1.m}\end{itemize}}

\numberwithin{equation}{chapter}
\begin{document}

\title{\centering Sloshing of viscous fluids:\\ Application to aerospace}
\author{Benjamin A. H. MEUNIER, Maxime C. N. ROUX\\}
\date{June 2021}

\begin{titlingpage} 
\begin{center}

\vspace*{4cm} 
\begin{LARGE}
\thetitle \\
\end{LARGE}
\vspace{0.5cm}
\theauthor
\vspace{0.5cm} 
\thedate
\vspace{2cm}
\begin{figure}[H]
    \centering
    \includegraphics[scale=0.9]{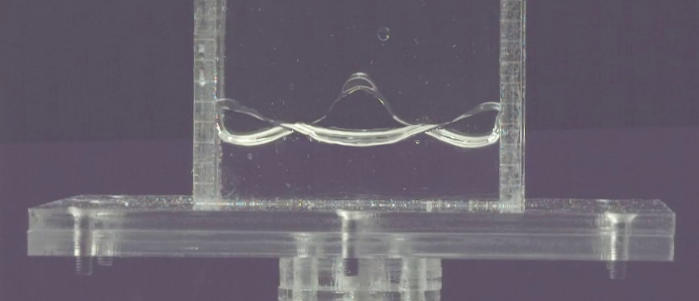}
\end{figure}
\begin{figure}[b!]
    \centering
    {\includegraphics[scale=0.11]{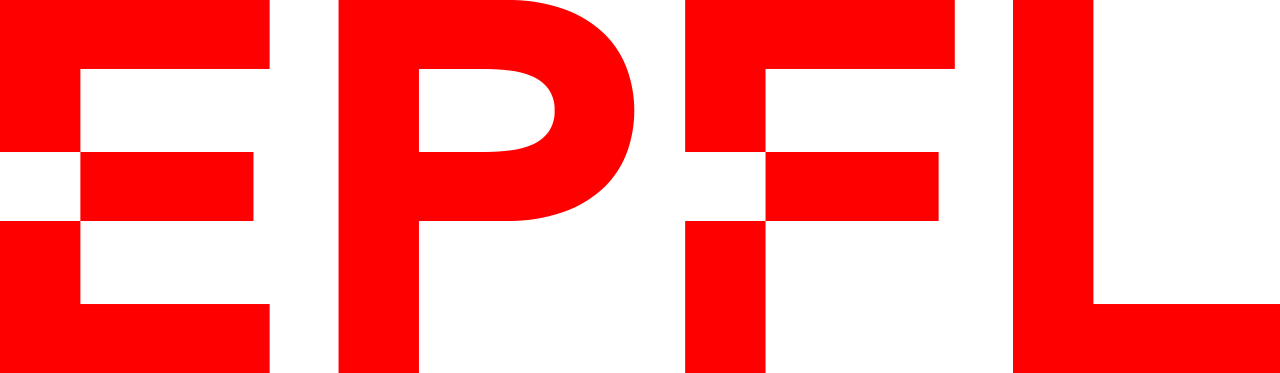}}
    \hfill
    {\includegraphics[scale=0.045]{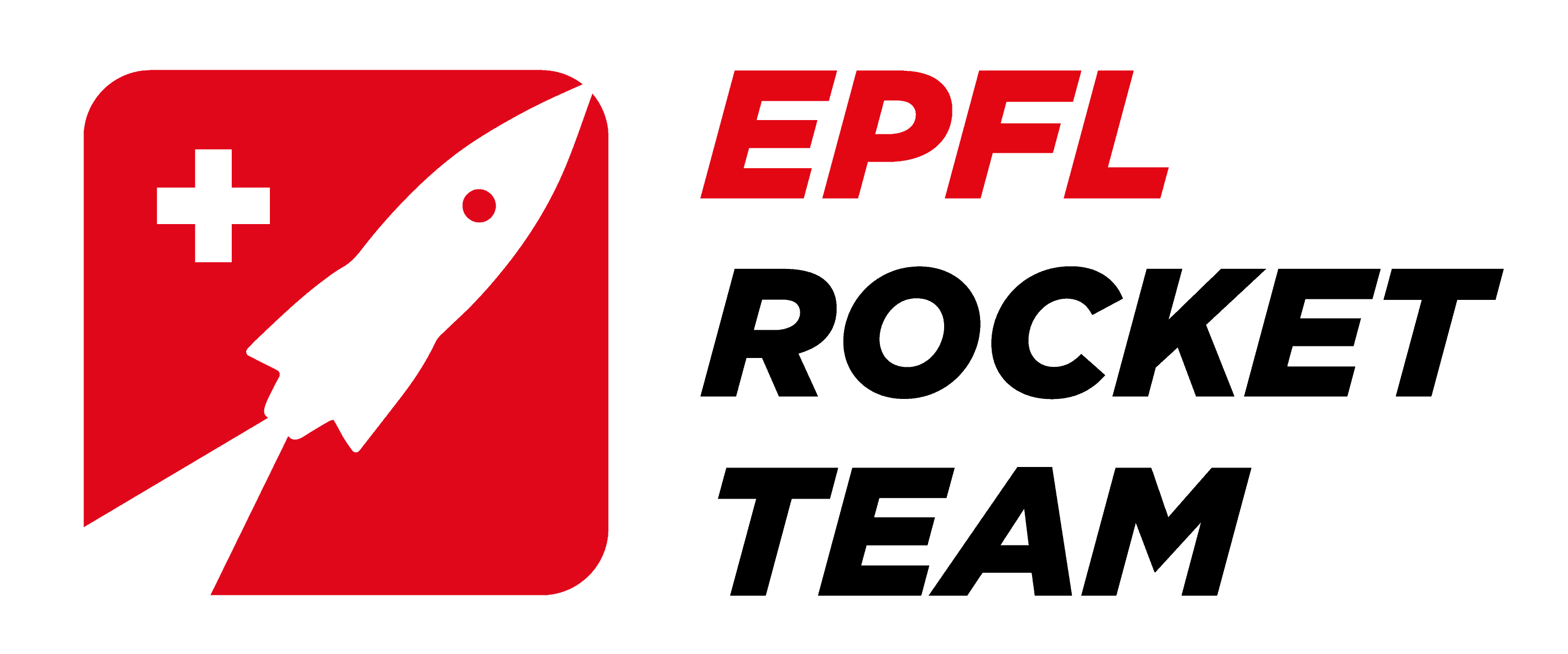}}
\end{figure}

\end{center}
\end{titlingpage}

\pagenumbering{Roman}

\tableofcontents

\renewcommand\thesection{\ifnum \value{chapter}>0  \thechapter.\fi\Roman{section}}

\newpage

\section{Acknowledgements}
This interdisciplinary project would not have been possible without the contribution and help of many people and organisations whom we gratefully acknowledge.\\

First of all, we would like to thank the teaching staff of the Practical Work section of the Department of Physics and in particular Dr. Daniele Mari for taking the time to guide us through our project. His advice and the discussions we had with him about the attenuation phenomena were very valuable.\\

We would also like to express our gratitude to the technical team of the Practical Work section of the Physics department for their practical advice and availability. Special thanks go to Antonio Gentile for his help in designing and setting up the experiments.\\

We would also like to thank the auditorium experiments team for agreeing to lend us their high-speed camera for one semester, a central element of our project, without which no experiment would have been possible.\\

We would also like to thank the entire EPFL Rocket Team for their hospitality and kindness. Two of its members were particularly involved throughout the project and we express our gratitude to them.\\

First of all, Loup Cordey, the payload manager and referent for our project within the association. His creativity inspired us both in the choice of our project and in its realisation. His experience in implementing projects in collaboration with the physics laboratories was of great help to us.
A big thank you also goes to Timour Jestin, who put us in touch with the association and helped us to set up several experiments during the project.\\

We would also like to thank Robopoly and especially its Vice-President, Amélie Senaud, for lending us their thermal camera.\\

Finally, we would like to express our gratitude to PolyLAN and in particular to Cédric Hölzl, Server Manager, member of the Intranet team, for allowing us access to their premises, providing us with powerful computers and helping us to configure them in order to carry out our numerical simulations.

\newpage

\section{Glossary}
\begin{itemize}
    \item R is the radius of the cylindrical tank
    \item $P$ is the pressure of the fluid
    \item $p_{ext}$ is the ambient pressure
    \item $p_{s}$ is the Laplacian pressure
    \item $\rho$ is the density of the fluid
    \item $\chi$ is the compressibility of the fluid
    \item $\eta$ is the dynamic viscosity coefficient of the fluid
    \item ${\eta}^{*}$ is the special viscosity coefficient of the fluid
    \item $\nu\equiv\eta/\rho$ is the kinematic viscosity coefficient of the fluid
    \item $\gamma$ is the surface tension of the fluid
    \item $\kappa$ is the curvature of the fluid surface
    \item $R_1$ is the radius of curvature of the fluid along a given axis
    \item $R_2$ is the radius of curvature of the fluid along an orthogonal axis to the preceding axis
    \item $\textbf{T}$ is the pseudo-vector of the fluid vortex
    \item $\textbf{u}$ is the velocity field of the fluid in the ground reference frame
    \item $\textbf{u}_{rel}$ is the velocity field of the fluid in the rocket reference frame
    \item $\textbf{a}$ is the force field per unit of force experienced by the fluid
    \item $g$ is the earth gravity
    \item $\Ddot{Z}_0$ is the vertical acceleration of the rocket
    \item $\textbf{V}_0$ is the speed of the rocket
    \item $h$ is the height of the liquid in the tank
    \item $\xi$ is the $z$ coordinate of the fluid surface
    \item $\Phi$ is the potential of the fluid in the ground reference frame
    \item $\Phi_0$ is the part of the fluid potential related to the movement of the rocket in the ground reference frame
    \item $\Tilde{\Phi}$ is the potential of the fluid in the rocket reference frame
    \item $J_m$ is the first order Bessel function of order $m$
    \item $\epsilon_{mn}$ is the n-th root of the derivative of $J_m$
    \item $\lambda_{mn}\equiv\epsilon_{mn}/R$ is the n-th root of the derivative of $J_m$ normalised by the radius
    \item $\omega_{mn}$ is one of the natural pulsations of the propellant in the tank
    \item $f_{mn}$ is one of the natural frequencies of the propellant in the tank
\end{itemize}

\newpage

\renewcommand\thesection{\arabic{section}}
\pagenumbering{arabic}

\chapter{Introduction}
   Sloshing of viscous liquids within launch vehicle tanks is a recurring problem when using liquid propellant engines. Many launchers, from the Soviet N-1 lunar rocket to the French Emerald series of launchers, have been lost due to sloshing. Two phenomena are at work.\\
    
    The first is called the "Pogo effect". This is a feedback phenomenon involving the propulsion, the structure and the liquids contained in the tanks of the launcher. During a flight, the sloshing of propellants in the tanks leads to pressure variations in the tanks. These pressure variations disrupt the power supply to the engine and cause variations in engine thrust. This causes vibrations in the launcher structures, which in turn causes the liquids in the tanks to slosh. If this phenomenon becomes positive, it can lead to damage or destruction of the launcher.\\
    The second phenomenon is the application of lateral forces and moments of force due to sloshing. This can lead to problems with the trajectory \cite{Farhat} and can result in the destruction of the launcher.\\
   Mitigating propellant sloshing in launch vehicles is therefore a critical challenge in the aerospace industry.\\
    
    This interdisciplinary project is carried out with the EPFL Rocket Team (ERT). The aim of this association is to build rockets that will fly in the Spaceport America Cup. This is a student competition that challenges students to send a rocket as close as possible to a given altitude. The integrity of the launcher and the stability of its trajectory are therefore essential. Our approach is part of the ERT's development of new liquid propellant engines and has a threefold objective.\\
    Firstly, the aim is to gather and create theoretical and technical bases on the subject that can subsequently be used by ERT engineers in the development of tanks and the choice of propulsion.\\
    The aim is also to develop innovative solutions based on the properties of the liquid in the tank in order to alleviate the problems associated with this phenomenon. As things stand at present, the various players in the field favour structural modifications to the launcher and the tank. Our project, on the other hand, will focus more on modifying the characteristics of the liquid by various means in order to reduce sloshing.\\
    Finally, it aims to gather relevant elements from a theoretical and experimental point of view in order to allow ERT engineers to develop an experiment for the Payload of one of their rockets.\\
    Our report will present three different, though complementary, lines of research. Firstly, an analytical, experimental and numerical approach to characterise the resonant frequencies of rocket tanks will be carried out. Secondly, we will study analytically and experimentally the attenuation of sloshing under free oscillations. Both single and two-phase viscous liquids will be studied. Thirdly, we will carry out an analytical and numerical study of the sloshing under forced oscillations.


\chapter{Resonant frequencies of liquids in rigid tanks}
This first part deals with the resonant frequencies of propellants in rocket tanks. They will first be determined analytically. Then, the theoretical model will be verified experimentally. Finally, a numerical implementation will be performed and used to determine these resonant frequencies for the rockets of the EPFL Rocket Team.

\section{Analytical approach}
The analytical approach is divided into three parts. The first is very general. It allows to establish a differential equation based on the characteristics of the liqui, but not of the tank. The other two will allow us to obtain the resonant frequencies in the case of rectangular tanks and then in the case of cylindrical tanks. This section and the related work are largely inspired by the work of R.A. Ibrahim \cite{Ibrahim}.
\subsection{General approach and fluid physics}
The liquid propellant contained in the tank of the EPFL Rocket Team is a Newtonian fluid. It is therefore described by the Navier-Stokes equation given by Equation \eqref{NavierStokesA} \cite{Meister}.

\begin{equation}
    \rho \textbf{ a} - \nabla P - \rho \nabla \frac{u^2}{2} + \eta {\nabla}^{2} \textbf{u} + (\eta+{\eta}^{*}) \nabla (\nabla \cdot \textbf{u}) = \rho  \textbf{ } \frac{\partial \textbf{u}}{\partial t} - 2 \textbf{ }\rho \textbf{ u } \wedge \textbf{ T}
\label{NavierStokesA}
\end{equation}
where $\textbf{T}=\frac{1}{2} \nabla \wedge \textbf{u}$. \\
The propellant also respects the continuity equation given by Equation \eqref{ContinuitéA} \cite{Meister}.
\begin{equation}
    \frac{\partial \rho}{\partial t}+ \nabla \cdot (\rho \textbf{u}) =0
\label{ContinuitéA}
\end{equation}
Assuming that propellants are perfect, incompressible and irrotational fluids, and following the reasoning described in Appendix A.\ref{1_1_1}, Equation \ref{NavierStokes7A} is obtained.
\begin{equation}
    g \xi + \frac{\gamma }{\rho}\left(\frac{1}{R_1} + \frac{1}{R_2}\right) + \frac{1}{2}{(\nabla \Tilde{\Phi} \cdot \nabla \Tilde{\Phi})} - \frac{\partial \Tilde{\Phi}}{\partial t} - \frac{\partial \Phi_0}{\partial t} - \frac{1}{2} {V_0}^2= 0
    \label{NavierStokes7A}
\end{equation}

\subsection{Rectangular tank}
Consider a rectangular tank of length $L$, width $l$ and such that the height of the fluid at rest is $h$.\\
Following the reasoning described in Appendix A.\ref{1_1_2}, the propellant resonant frequencies given by Equation \eqref{fréquenceRA} are obtained.
\begin{equation}
    f_{mn} = \frac{1}{2 \pi}\sqrt{\left[(g + \Ddot{Z_0}) k_{mn} + \frac{\gamma}{\rho} {k_{mn}}^{3} \right] \tanh\left(k_{mn} h\right)}
    \label{fréquenceRA}
\end{equation}
where $k_{mn} = \pi \sqrt{(2m)^2/L^2 + (2n)^2/l^2}$.

\subsection{Cylindrical tank}
Consider a cylindrical tank of radius $R$ and such that the height of the fluid at rest is $h$.
    Following the reasoning detailed in Appendix  A.\ref{1_1_3}, the propellant resonant frequencies given by Equation \eqref{fréquenceCA} are obtained.
\begin{equation}
    f_{mn} = \frac{1}{2 \pi}\sqrt{\left[(g + \Ddot{Z_0})\lambda_{mn} + \frac{\gamma}{\rho } {\lambda_{mn}}^{3} \right] \tanh\left( {\lambda_{mn}h}\right)}
    \label{fréquenceCA}
\end{equation}
where $\lambda_{mn} = \varepsilon_{mn} / R$.\\

\section{Experimental approach}
Given the difficulty of establishing a rigorous experimental protocol to verify the Equation \eqref{fréquenceCA}, only Equation \eqref{fréquenceRA} which gives the resonant frequencies for a rectangular tank will be experimentally tested in this project. The Equations \eqref{fréquenceCA} and \eqref{fréquenceRA} are very similar, the only term that changes is a term that depends solely on the geometry of the tank. It will therefore be assumed that if Equation \eqref{fréquenceRA} giving the resonant frequencies for a rectangular tank is verified experimentally, Equation \eqref{fréquenceCA}, which gives the resonant frequencies for a cylindrical tank, will be considered valid in the context of this project.\\

\subsection{Setup and experimental protocol}
\label{montage}
\begin{figure}[H]
    \centering
    \includegraphics[scale=0.4]{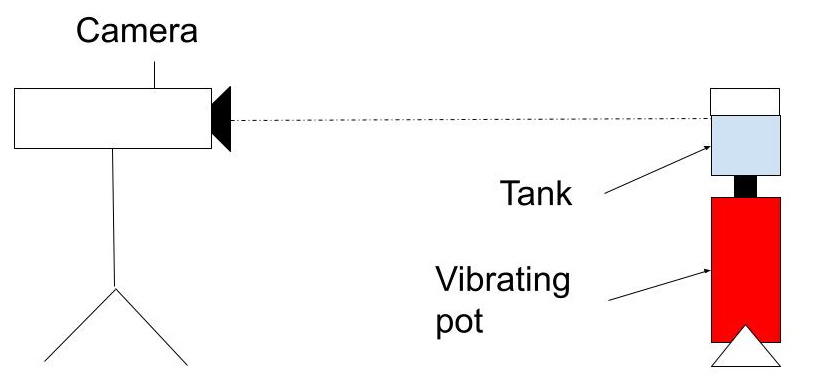}
    \caption{Experimental setup}
    \label{experience}
\end{figure}

\paragraph{Experimental setup}
The experimental setup used to verify Equation \eqref{fréquenceRA} is illustrated in Figure \ref{experience}.
It consists of two parts. The first part consists of a rectangular tank with the dimensions $L = 6.9$ cm and $l = 1.5$ cm, which is attached to a vibrating container which is controlled by a function generator connected to an amplifier. The generator is set to the sinusoidal mode, which produces a sinusoidal signal of selected amplitude and frequency. The second part consists of a Phantom Research v411 camera which is connected to a computer and set to record 300 frames per second in HD (1280 x 800). An LED light is used to illuminate the tank to ensure good video quality.

\paragraph{Preliminary observations}
Using the setup presented above, preliminary experimental observations are collected which will allow us to establish a precise protocol are collected.\\
The waveform associated with a given mode is generally obtained around the theoretical frequency with a fairly good accuracy  ($\pm 0.5$ Hz). By moving too far away from the model frequency, the waveform may remain that of the predicted mode, but there will be a spatial phase shift.
Furthermore, the first two modes (1.0) and (2.0) as well as the modes with frequencies close to each other are difficult to excite. At the same time, the high modes are difficult to distinguish visually from each other.
In addition, when the liquid level is very low, it interacts too much with the walls and the bottom of the tank to obtain usable results. These effects are not taken into account in the analytical reasoning.
Finally, the higher the liquid level, the more difficult it is to excite the resonance in a steady state.

\paragraph{Experimental protocol}
A frequency sweep is performed around a resonant frequency associated with the $(3,0)$ mode with a step size of $0.03$ Hz for a given excitation height and amplitude. This mode is chosen for its simplicity of excitation and purity to ensure accurate results. A film is taken with the camera between the 5th and 6th second after the start of the excitation. The maximum liquid height is obtained in this time interval and measured with the image processing software \textit{paint.net} by counting the number of vertical pixels between the liquid level at rest and the highest point of the wave. The experimental resonant frequency will be the one associated with the highest maximum height. This protocol allows a more rigorous experimental verification than a simple observation of the wave shape. By convention,  $h_{max}$ is measured by choosing the vertical position of the free surface at rest as the origin of the heights.\\
In parallel, the temperature of the liquid will be measured to determine its density and surface tension. With these two parameters, it will be possible to use Equation \eqref{fréquenceRA} to calculate the theoretical resonant frequency and to compare it with the experimental frequency obtained.  
\subsection{Observations and experimental results}
The following experiments are carried out using demineralised water.

\begin{figure}[H]
    \centering
    \includegraphics[scale=1]{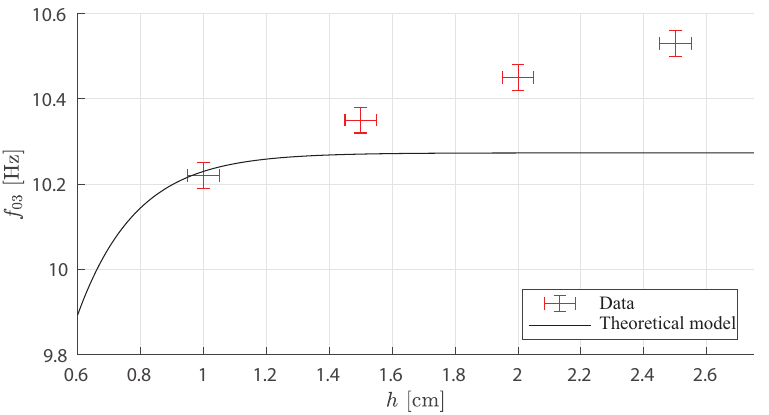}
    \caption{Comparison between the theoretical and experimental frequency of the $(3,0)$ mode as a function of water level height.}
    \label{water_30}
\end{figure}

\paragraph{Findings}
 The protocol is carried out at heights $h$ of $1$ cm, $1.5$ cm, $2$ cm and $2.5$ cm.\\
For $h=2$ cm,  the results of the frequency sweep are shown in Figure \ref{water_30_h_2_0}. For other heights, the results are given in Appendix A.\ref{1_2_2}.
\begin{figure}[H]
    \centering
    \includegraphics[scale=1]{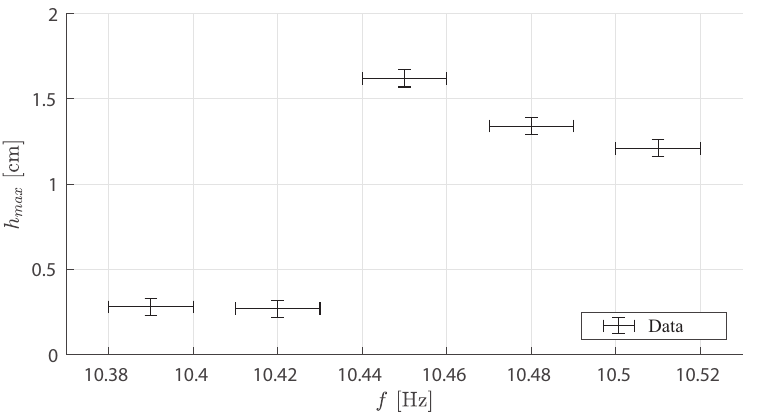}
    \caption{Frequency sweep for the  $(3,0)$ mode at a height of $h = 2$ cm.}
    \label{water_30_h_2_0}
\end{figure}
Using a thermal camera, the temperature of the water is measured at $21.3^{\rm{o}}$C (Appendix A.\ref{1_2_1}). The values in the tables are taken for pure water at $20.0^{\rm{o}}$C (Appendix A.\ref{2_0_0}).\\
Figure \ref{water_30} shows the experimental frequencies obtained for different heights, and compares them with those predicted by Equation \eqref{fréquenceRA}.

\subsection{Discussion}
The measured frequencies are higher than the theoretical ones. This can be explained by the fact that the theoretical frequency is obtained for a perfect fluid and not for a viscous fluid. In addition, edge effects that are largely due to the surface tension between the liquid and the container are observed in the images in Appendix  A.\ref{1_2_1}. These effects are neglected in the considerations leading to the analytical expressions for the resonant frequencies and are dependent on the material used for the tank. However, the results remain relatively close and this model will therefore be used thereafter.
\section{Numerical application to the rockets of the EPFL Rocket Team}
\begin{figure}[H]
    \centering
    \includegraphics[scale=1]{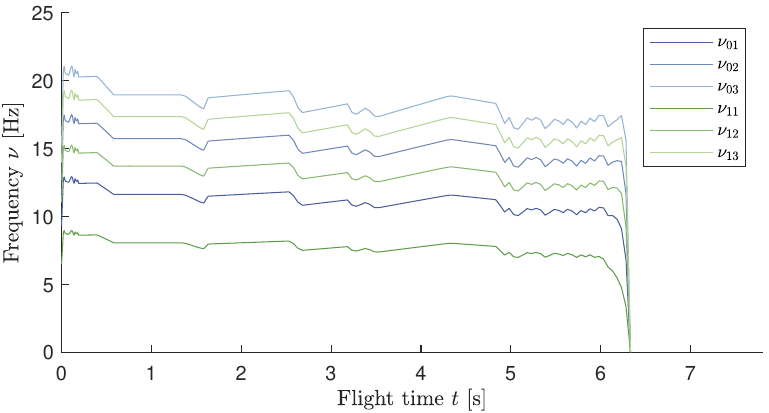}
    \caption{Evolution of the first resonant frequencies during the propulsion phase.}
    \label{f}
\end{figure}

The objective of this third part will be to write a Matlab script in order to calculate the different resonant frequencies of the liquids contained in the tanks of the rockets of the EPFL Rocket Team..\\
The scripts used to solve the Equation \eqref{fréquenceCA} throughout the flight are given in Appendix A.\ref{1_3_1}. Using the propellant data, tank drawings and previous flight data given in Appendix \ref{3_0_0}, the evolution of the first resonant frequencies of the Rocket Team Bella-Lui rocket is determined. They are shown in Figure \ref{f}. The choice of these first frequencies is purely arbitrary and follows the indications of the ERT teams who informed us that the rocket was mainly subjected to excitations at frequencies below about twenty hertz. Experiments will be carried out on this subject during future launches, and experimental data can then be collected to test the model. 

\section{Conclusion for resonant frequencies}
Thus, despite differences of up to $0.15$ Hz between the model and the theory, which can be explained by the approximations made to develop the theoretical model, the Equation \eqref{fréquenceRA} allows us to describe correctly the resonant frequencies in a rectangular tank. In view of the similarity with the cylindrical case, which differs only in the change of the tank geometry, the Equation \eqref{fréquenceCA} in the cylindrical case is considered in this project to be correct to describe the resonant frequencies in a cylindrical tank.\\
Algorithms have been implemented to predict the resonant frequencies of the Bella-Lui rocket fuel. The first resonant frequencies are between 10 and 20 Hz. To reduce the risk to the rocket, ERT has two options.  It is possible to modify the characteristics of the liquid, either to increase the resonant frequencies by increasing the $\gamma/\rho$ ratio, or to attenuate the wave height by increasing the sloshing attenuation coefficient of the liquid.. It is this second option that will be studied in the second part.

\chapter{Mitigation of free sloshing of a single and two phase liquid in a rigid tank}
This second part will focus on the attenuation of free sloshing of a single and two phase liquid. The first part of this approach will be analytical and based on energy dissipation, the second will be experimental.
\section{Analytical approach}
First a global approach based on energy dissipation will be considered, then the attenuation coefficient for the sloshing of single phase viscous fluids will be established in the rectangular and cylindrical cases.
\subsection{Energy dissipation}
In the case of free oscillations of a viscous fluid, the height of the free surface of the fluid is given by Equation \eqref{ansatzKi} \cite{Bauer1997}.
\begin{equation}
\xi(x,y,t) = \mathrm{Re}\left[\sum_{m,n} {\xi}_{mn}(x,y,t)\right]
\label{ansatzKi}
\end{equation}
where ${\xi}_{mn}(x,y,t)$ is the vertical coordinate of the waves at the surface due to the eigenmode $(m,n)$. The free oscillations of a viscous fluid are damped. The time dependence of ${\xi}_{mn}(x,y,t)$ is assumed to be described by Equation \eqref{ansatzKimn} \cite{Ibrahim}.
\begin{equation}
    {\xi}_{mn}(x,y,t) = {\xi}_{mn}^*(x,y) e^{-(i\omega_{mn}+\alpha_{mn})t}
\label{ansatzKimn}
\end{equation}
where $\alpha_{mn}=1/\tau_{mn}$ is the damping coefficient of the eigenmode $(m,n)$ and $\tau_{mn}$ is the characteristic time of the attenuation.
By combining the Equation \eqref{ansatzKi} and Equation \eqref{ansatzKimn}, the Equation \eqref{Ki} is obtained.
\begin{equation}
\xi(x,y,t) = \sum_{m,n} \mathrm{Re}[{\xi}_{mn}^*(x,y)e^{-(i\omega_{mn}+\alpha)t}]
\label{Ki}
\end{equation}
The kinetic energy of the fluid is given by Equation
\eqref{energieCinetique}.
\begin{equation}
    T_{mn}=\frac{1}{2}\rho\int_V \mathbf{u_{mn}}^2\dd V
\label{energieCinetique}
\end{equation}
where $\mathbf{u_{mn}}$  is the velocity of the fluid according to the Lagrangian description. The potential energy of the fluid is given by Equation \eqref{energiePotentielle}.

\begin{equation}
    \Pi_{mn}=\frac{1}{2}\rho g \xi_{mn}^2
\label{energiePotentielle}
\end{equation}

\begin{equation}
    \xi_{mn}(t)=\xi_{mn}^* e^{-(i\omega_{mn}+\alpha_{mn})t}
\label{ansatzQmn}
\end{equation}
The average energy $\langle E\rangle (t)$ is therefore given by Equation \eqref{energieMoyenne}.

\begin{equation}
    \langle E_{mn}\rangle=\langle T_{mn}+\Pi_{mn}\rangle= \rho g {\xi_{mn}^{*}}^2e^{-\alpha_{mn} t}
\label{energieMoyenne}
\end{equation}
The damping coefficient $\alpha_{mn}$ is therefore given by Equation \eqref{alpha}.
\begin{equation}
    \alpha_{mn}=\frac{\langle |\Dot{E}_{mn}|\rangle}{2\langle E_{mn}\rangle}
\label{alpha}
\end{equation}
Moreover, in a liquid, the Equations \eqref{EnergiePointMoy} and \eqref{EnergieLiquide} are verified \cite{Laudau}. They allow analytical predictions of the coefficients which will be presented later.
\begin{equation}
    \langle|\Dot{E}|\rangle = \frac{1}{2} \rho \int {\(\nabla \mathbf{u}+\nabla \mathbf{u}^{\top}\)}^2 dV
\label{EnergiePointMoy}
\end{equation}
\begin{equation}
    E = \frac{1}{2} \rho \int \mathbf{u}^2 dV
    \label{EnergieLiquide}
\end{equation}
The energy dissipation can be decomposed on the one hand into a dissipation at the free surface associated with $\alpha_{fs}$, and on the other hand into a dissipation through the solid walls at the sides and bottom of the tank associated with a damping coefficient $\alpha_{wl}$. The total dissipation coefficient is the sum of  $\alpha_{fs}$ and $\alpha_{wl}$.

\subsection{Rectangular tank}
For a rectangular tank, the damping coefficients are given by Equations \eqref{alphaFSRectangle} and \eqref{alphaWLRectangle} \cite{Sauret2015}.

\begin{equation}
    \alpha_{fs} = M\nu k_{mn}^2
\label{alphaFSRectangle}
\end{equation}

\begin{equation}
    \alpha_{wl} = N\sqrt{\omega_{mn}\nu}\left(\frac{1}{l}+\frac{1}{L}\right)
\label{alphaWLRectangle}
\end{equation}
where $M$ and $N$ are proportionality coefficients to be determined experimentally.

\subsection{Cylindrical tank}
For a cylindrical tank, the damping coefficients are given by Equations  \eqref{alphaFSCylindre} and \eqref{alphaWLCylindre} \cite{Ibrahim}.

\begin{equation}
    \alpha_{fs}=2\nu\lambda_{mn}^2
\label{alphaFSCylindre}
\end{equation}

\begin{equation}
    \alpha_{wl}=\frac{1}{2R}\sqrt{\frac{\nu\omega_{mn}}{2}}\frac{1+(n/\lambda_{mn}R)^2}{1-(n/\lambda_{mn}R)^2}
\label{alphaWLCylindre}
\end{equation}

\section{Experimental approach}
\subsection{Experimental setup and protocols}
\paragraph{Attenuation coefficient}
The experimental setup used is exactly the same as the one used for the resonant frequencies.\\
For a given liquid height, the liquid is excited to the desired mode. The frequency associated with the mode used in the experiment is that given by Equation \ref{fréquenceRA}.  When the liquid has reached a steady state, the function generator is switched off. A film is then taken with the camera. The evolution of the height of the free surface at a fixed horizontal position is studied. Since the film has finite resolution, the horizontal position of this point is chosen at a vibrational belly (hence a pressure node). Equation \eqref{ansatzKimn} predicts that the height versus time has an exponentially damped sinusoidal profile. The damping coefficient corresponds to the coefficient $\alpha=\alpha_{fs}+\alpha_{wl}$ studied in the previous section. For reasons of time, only the local minima will be sampled, i.e. the points where $\sin(-\omega t)=1$ is verified.  The neperian logarithm is then applied to the data in order to perform a linear regression and thus obtain the experimental coefficient $\alpha_{exp}$.

\paragraph{Coefficients M et N}
The protocol for the determination of the total attenuation coefficient $\alpha$ is performed for the modes $(2,0)$, $(3,0)$, $(4,0)$, $(4,1)$ et $(5,1)$  with a liquid height at rest of $h = 5$ cm. A linear regression is then performed to determine the coefficients $N$ and $M$ of Equation \eqref{alphareg}.
\begin{equation}
    \alpha = M \alpha_{fs} + N \alpha_{wl}
\label{alphareg}
\end{equation}
where $\alpha_{fs}$ is given by Equation \eqref{alphaFSRectangle} and $\alpha_{wl}$ is given by Equation \eqref{alphaWLRectangle}.

\paragraph{Two-phase liquid}
For qualitative purposes, the protocol for finding the total attenuation coefficient $\alpha$ is carried out for a two-phase liquid with a fixed height at rest of $h = 5$ cm.

\subsection{Results for a single-phase liquid}
\begin{figure}[H]
    \centering
    \includegraphics[scale=1]{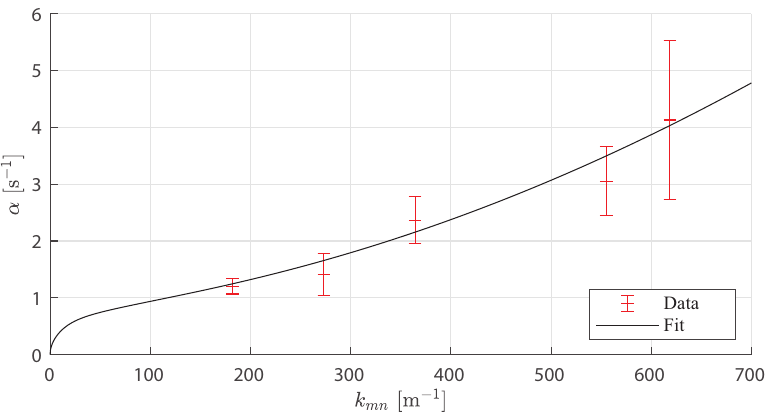}
    \caption{Experimental attenuation as a function of mode and with a fit presented in Equation \eqref{alphareg}.}
    \label{alpha_kmn}
\end{figure}

The liquids in the tank are measured at $21.3^{\rm{o}}$C using a thermal camera  (Appendix A.\ref{1_2_1}). The characteristics of the liquid will therefore be taken at  $20.0^{\rm{o}}$C in Appendix C.\ref{2_0_0}.
\paragraph{Swap mode}
The protocol is performed with demineralised water. The results and the regression are presented in Figure \ref{alpha_kmn}. The values obtained are $M = 5.5$ and $N = 1.8$.

\subsection{Results for a two-phase liquid}
\begin{figure}[H]
    \centering
    \includegraphics[scale=1]{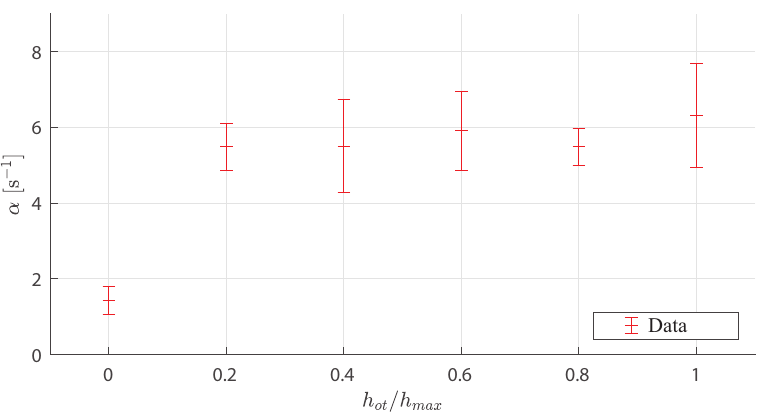}
    \caption{Experimental attenuation of a two-phase liquid as a function of oil proportion}
    \label{water_oil}
\end{figure}

\paragraph{Liquid with a water phase and an oil phase}
The protocol is applied with a liquid consisting of a demineralised water phase and a sunflower oil phase. The attenuation coefficient as a function of the oil fraction in the liquid is shown in Figure  \ref{water_oil}.

\begin{figure}[H]
    \centering
    \includegraphics[scale=0.5]{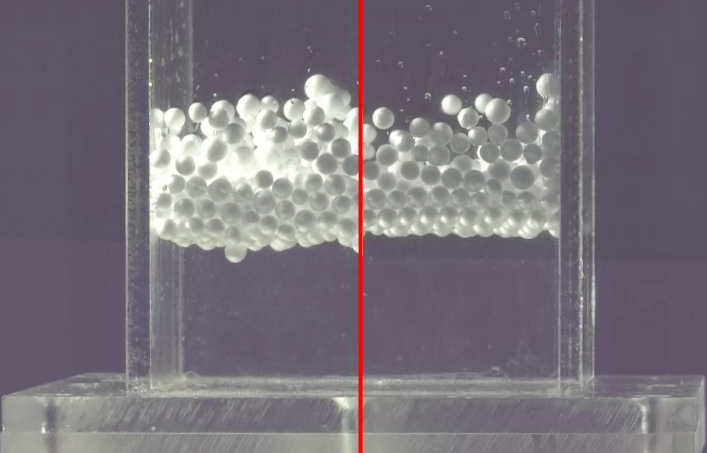}
    \caption{Maximum excitation (left) and minimum excitation (right) of height under forced oscillations for a square tank of side $L = 10$ cm.}
    \label{carrepoly}
\end{figure}

\begin{figure}[H]
    \centering
    \includegraphics[scale=0.5]{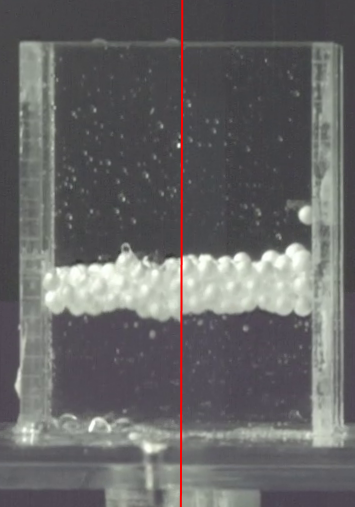}
    \caption{Maximum excitation (left) and minimum excitation (right) of height under forced oscillations for a rectangular side tank $L = 6.9$ cm et $l = 1.5$ cm.}
    \label{rectanglepoly}
\end{figure}

\paragraph{Polystyrene beads and water}
An attempt was made to perform the same experiment with a layer of polystyrene beads with a diameter between $d = 2$ and $d= 4$ mm. However, as shown in Figures \ref{carrepoly} and \ref{rectanglepoly}, it is very difficult if not impossible to excite the liquid in the tank in this case.

\subsection{Discussion}
\paragraph{Swap on modes} The regression performed on Figure \ref{alpha_kmn} fits the data well. The data collected do therefore not allow this model to be rejected. However, two parameters are used for a fit on only five points. Moreover, the analytical reasoning leading to an analytical expression of $\alpha$ in the rectangular case repeatedly uses dimensional analysis to advance the reasoning \cite{Bronfort}. 
This reasoning is not very robust; therefore numerous additional robust experimental data need to be collected in order to confirm or refute this model. To test it more reliably, rectangular containers of different sizes and fluids of different viscosities should be used.\\
In addition, capillary effects have been taken into account to develop the Equation \eqref{alphaWLRectangle}. As this equation does not show an explicit dependency on the surface tension of the liquid, the coefficient $N$ must therefore depend on the surface tension between the liquid and the wall material. Further analysis is required to obtain an explicit dependency on the material and the surface tension of the liquid.

\paragraph{Liquid with a water phase and an oil phase} Figure \ref{water_oil}  shows that the attenuation is the same for 20\% oil and for pure sunflower oil. This result suggests that the liquid at the surface (generally the least dense) has the greatest influence on the attenuation. This result can be explained by the fact that the depth of the waves, i.e. the amplitude of the time sinusoid in a fixed horizontal coordinate, is quite small compared to the height of the container.\\
Furthermore, it is the liquid at the surface that determines the surface tension of the two-phase liquid, which plays a crucial role in attenuation. Since the surface tension of sunflower oil is twice that of water  (Appendix C.\ref{2_0_0}), attenuation is strongly affected by this factor.\\
Furthermore, the attenuation when using sunflower oil is more than three times higher than that of pure water. This result suggests that a thin layer of a high viscosity, low density liquid on the surface can significantly reduce sloshing and attenuate resonance.

\paragraph{Polystyrene beads and water} Figures \ref{carrepoly} and \ref{rectanglepoly} show that at the minimum and maximum of the excitation the bead block remains in one block, no waves are observed. No stationary mode corresponding to a resonance could be excited with the experimental protocol used, for frequencies between 2 and 40 Hz. The beads form a compact block which is held together and immobile relative to the tank by static friction between the beads and with the walls. The low density of the beads allows them to concentrate on the surface and thus maximise the static friction between them which increases their effective viscosity.\\
These results suggest that beads can be very effective in preventing resonant modes from being excited. Polystyrene beads were chosen because of their low density and ease of procurement. More effective beads in this role will be sought and investigated in further research.\\
It is also possible that beads on the surface break the waves of the resonant modes when excited. This hypothesised effect could not be tested due to the impossibility of forming waves on the surface with this experimental protocol.

\section{Conclusion for the attenuation of free oscillations}
Thus, despite a weak theoretical model and results that are difficult to analyse from a quantitative point of view, qualitative conclusions can be drawn. The higher the coefficient $k_{mn}$ associated with the $(m,n)$ mode, the higher the attenuation associated with this mode. Adding a layer of a more viscous liquid, even a thin one, will also significantly increase the attenuation coefficient. Finally, adding a layer of polystyrene beads seems to prevent excitation. However, experiments are still needed to find out exactly which physical phenomena explain this.

\chapter{Forced oscillations of viscous liquids in a rigid tank}
This section is devoted to the study of the sloshing of a liquid in a vertical cylindrical tank exposed to a sinusoidal excitation along a horizontal axis. In flight, this excitation corresponds to structural vibrations of the rocket caused by friction with the air.\\
\section{Analytical approach}
Consider a vertical cylinder of radius $R$ containing an incompressible viscous fluid at rest of height $h$. It is exposed to an excitation such that the horizontal position (measured along a Cartesian axis $\hat{\mathbf{x}}$) of its principal axis is described by the Equation  \eqref{excitationSin}.
\begin{equation}
    x(t)=X_0\sin(\Omega t)
\label{excitationSin}
\end{equation}
 where $X_0$ is the amplitude of the excitation and $\Omega$ est sais its pulsation. Maximum force $F_{x,\max}=\max\left(\mathbf{F}\cdot\hat{\mathbf{x}}\right)$ is given by Equation \eqref{FxMax}, where $\mathbf{F}$ is the force exerted by the fluid on the tank \cite{BauerForcee}.

\begin{equation}
\begin{aligned}
    F_{x,\max}&=m\Omega^2X_0+R\pi\eta\sum_{n=1}^{\infty}A_n\sinh\left(\xi_{n}h/R\right)\bigg\{(1+3\beta_n)\frac{\sqrt{\xi_n^2-i\Omega^*}}{2\xi_n}\\
   &\times J_1\left(\sqrt{\xi_n^2-i\Omega^*}\right)+\left(i\Omega^*-2\xi_n^2\right)\gamma_nJ_1(\xi_n)/\xi_n\bigg\}
\end{aligned}
\label{FxMax}
\end{equation}

The maximum vertical disturbance of the liquid $h_{max}$ is given by Equation \eqref{hMax} \cite{BauerForcee}.
\begin{equation}
\begin{aligned}
    h_{max}&=\max\bigg(\frac{1}{2i\Omega}\sum_{n=1}^{\infty}A_n\sinh\left(\xi_{n}h/R\right)\bigg\{\frac{\sqrt{\xi_n^2-i\Omega^*}}{2\xi_n}(1+\beta_n)\\
    &\times J_1\left(\sqrt{\xi_n^2-i\Omega^*}r/R\right)-2\xi_n\gamma_nJ_1(\xi_n r/R)\bigg\}\bigg)
\end{aligned}
\label{hMax}
\end{equation}

The $\xi_n$ are defined as the solutions of the Equation \eqref{xin}.
\begin{equation}
\begin{vmatrix}
    J_2\left(\sqrt{\xi^2-i\Omega^*}\right) & J_0\left(\sqrt{\xi^2-i\Omega^*}\right) & -\xi J_1'(\xi) \\ 
    J_2\left(\sqrt{\xi^2-i\Omega^*}\right) & -J_0\left(\sqrt{\xi^2-i\Omega^*}\right) & J_1(\xi) \\ 
    \sqrt{\xi^2-i\Omega^*}J_1\left(\sqrt{\xi^2-i\Omega^*}\right) & -\sqrt{\xi^2-i\Omega^*}J_1\left(\sqrt{\xi^2-i\Omega^*}\right) & \xi^2 J_1(\xi) 
\end{vmatrix}=0
\label{xin}
\end{equation}
Since this equation is a priori not analytically solvable, the reasoning presented thereafter uses numerical analysis methods.\\
The $A_n$  are defined as the linear coefficients of the system of equations  \eqref{An}.

\begin{equation}
\begin{cases}
    \sum_{n=1}^{\infty}A_n\xi_n\sinh(\xi_nh/R)&\bigg\{\frac{\xi_n^2-i\Omega^*}{\xi_n^2}(1+\beta_n)J_1'\left(\sqrt{\xi_n^2-i\Omega^*}(r/R)\right)\\
    &-J_2\left(\sqrt{\xi_n^2-i\Omega^*}(r/R)\right)+\beta_nJ_0\left(\sqrt{\xi_n^2-i\Omega^*}(r/R)\right)\\
    &-4\xi_n\gamma_nJ_1'(\xi_nr/R)\bigg\}=0\\
    \sum_{n=1}^{\infty}A_n\xi_n\sinh(\xi_nh/R)&\bigg\{J_2\left(\sqrt{\xi_n^2-i\Omega^*}(r/R)\right)+\beta_nJ_0\left(\sqrt{\xi_n^2-i\Omega^*}(r/R)\right)\\
    &+\frac{\sqrt{\xi_n^2-i\Omega^*}}{\xi_n^2r/R}(1+\beta_n)J_1\left(\sqrt{\xi_n^2-i\Omega^*}(r/R)\right)\\
    &-\frac{4\gamma_n}{(r/R)}J_1(\xi_nr/R)\bigg\}=0\\
    \sum_{n=1}^{\infty}A_{n}\frac{R}{\eta}\cosh(\xi_nh/R)&\bigg\{\frac{\sqrt{\xi_n^2-i\Omega^*}}{\xi_n}(1+\beta_n)\bigg[i\Omega^*\xi_n +\frac{1}{2}\sigma^*(\xi_n^2-i\Omega^*+\alpha^2)\tanh(\xi_nh/R)\bigg]\\
    &\times J_1\left(\sqrt{\xi_n^2-i\Omega^*}(r/R)\right)
    -\bigg[i\Omega^*(i\Omega^*+2\xi_n^2)\\
    &+\xi_n\sigma^*(\xi_n^2+\alpha^2)\tanh(\xi_nh/R) \bigg]\gamma_nJ_1(\xi_nr/R)\bigg\}=i\Omega^{*3}\frac{X_0}{R}\frac{r}{R}
\end{cases}
\label{An}
\end{equation}
where $r$ is the radial position, $\alpha^2=\rho gR^2/\sigma$ is the Bond number (representing the ratio between gravitation and the liquid-air surface tension), $\sigma^*=\sigma R/(\rho\nu^2)$ is a surface tension-viscosity parameter, and the $\beta_n$ and the $\gamma_n$ are defined by Equation \eqref{betaGamma}.

\begin{equation}
    \beta_n=\frac{J_2\left(\sqrt{\xi_n^2-i\Omega^*}\right)}{\xi_nJ_2(\xi_n)},\quad \gamma_n=\frac{J_2\left(\sqrt{\xi_n^2-i\Omega^*}\right)J_0(\xi_n)}{J_0\left(\sqrt{\xi_n^2-i\Omega^*}\right)J_2(\xi_n)}
\label{betaGamma}
\end{equation}

\section{Numerical approach}
To evaluate the equations for $h_{max}$ and for $F_{x,max}$, it is therefore necessary to determine the $\xi_n$ and the $A_n$.  This is a priori not possible analytically. Therefore, numerical analysis methods have to be found to obtain results.
\subsection{Solving the equation for $\xi_n$}

Let $f(x)$ be a meromorphic complex function (analytical except for a countable finite set of isolated points which are poles), whic is defined on a simply connected open $U\subset \C$ (any yaw is null-homotopic). Then the argument principle, which is a corollary of the residue theorem, corresponds to Equation \eqref{principeArgument}.
\begin{equation}
    \sum_{z_j\in F}v_{z_j}(f)\mathrm{Ind}_\gamma(z_j)=\frac{1}{2\pi i}\bigintssss_{\gamma\equiv\partial U}\frac{f'(z)}{f(z)}\mathrm{d} z
\label{principeArgument}
\end{equation}
where $F$ is the set of zeros and poles of $f$ on $U$,  $\mathrm{Ind}_\gamma(z)$ is the winding number of $z$ with respect to $\gamma$ (the number of revolutions $\gamma$ makes around $z$, traversed in the trigonometric direction), and $v_{z}(f)$ is the valuation of $f$ in $z$.  The characteristic equation \eqref{xin} can be written compactly as $g(\xi)=0$. $g$ does not contain a pole on any finite subset of $\C$ as none of the functions that appear in the equation diverge into a finite $z\in\C$. It is further assumed that all zeros of $g$ are of order 1, i.e. $v_{z_j}(g)=1\ \forall z_j\in F$. Equation \eqref{principeArgument} can therefore be used to determine the $\xi_n$.

\begin{figure}[H]
    \centering
    \subcaptionbox{}
    {\includegraphics[scale=0.11]{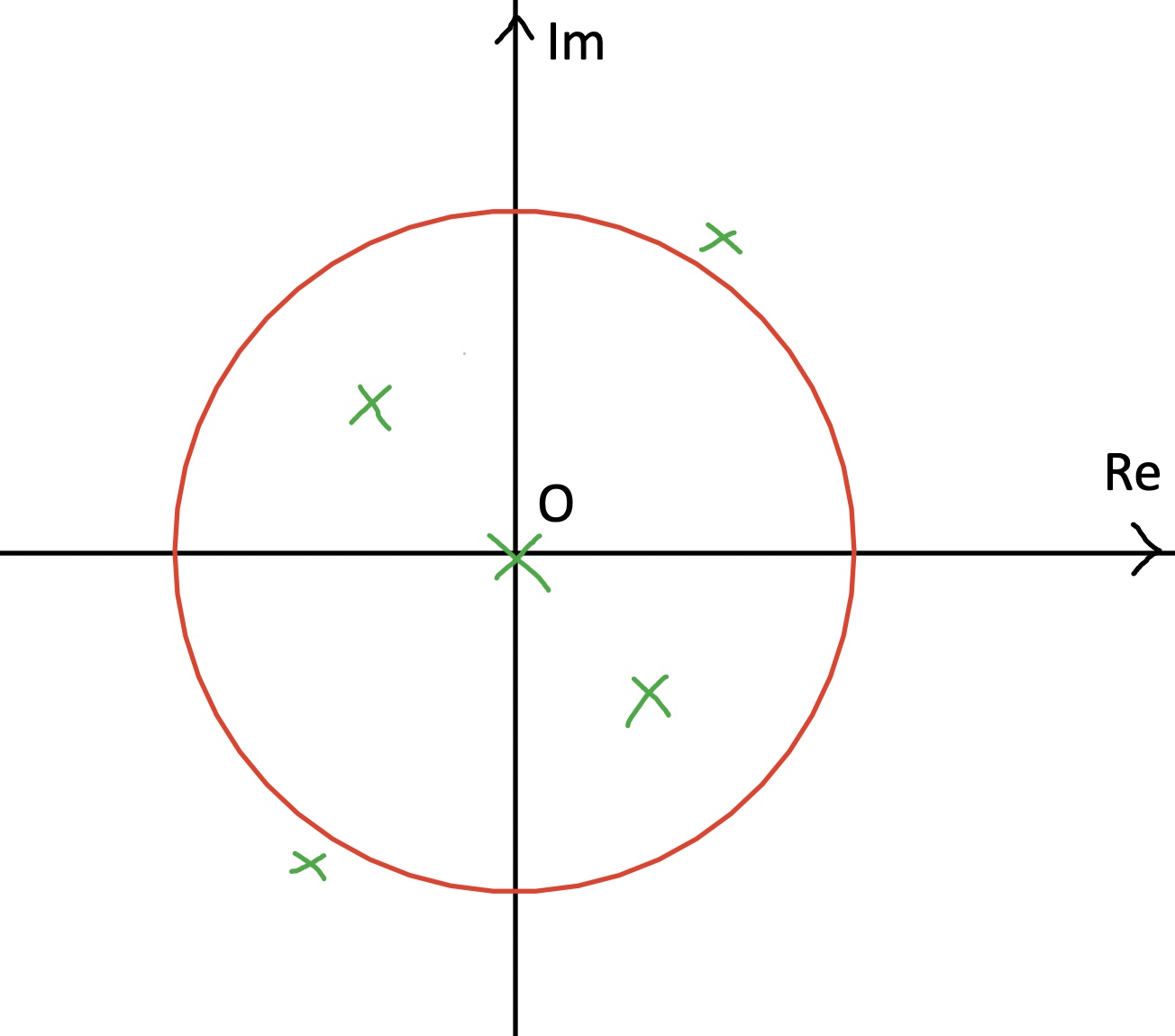}}
    \hfill
    \subcaptionbox{}
    {\includegraphics[scale=0.11]{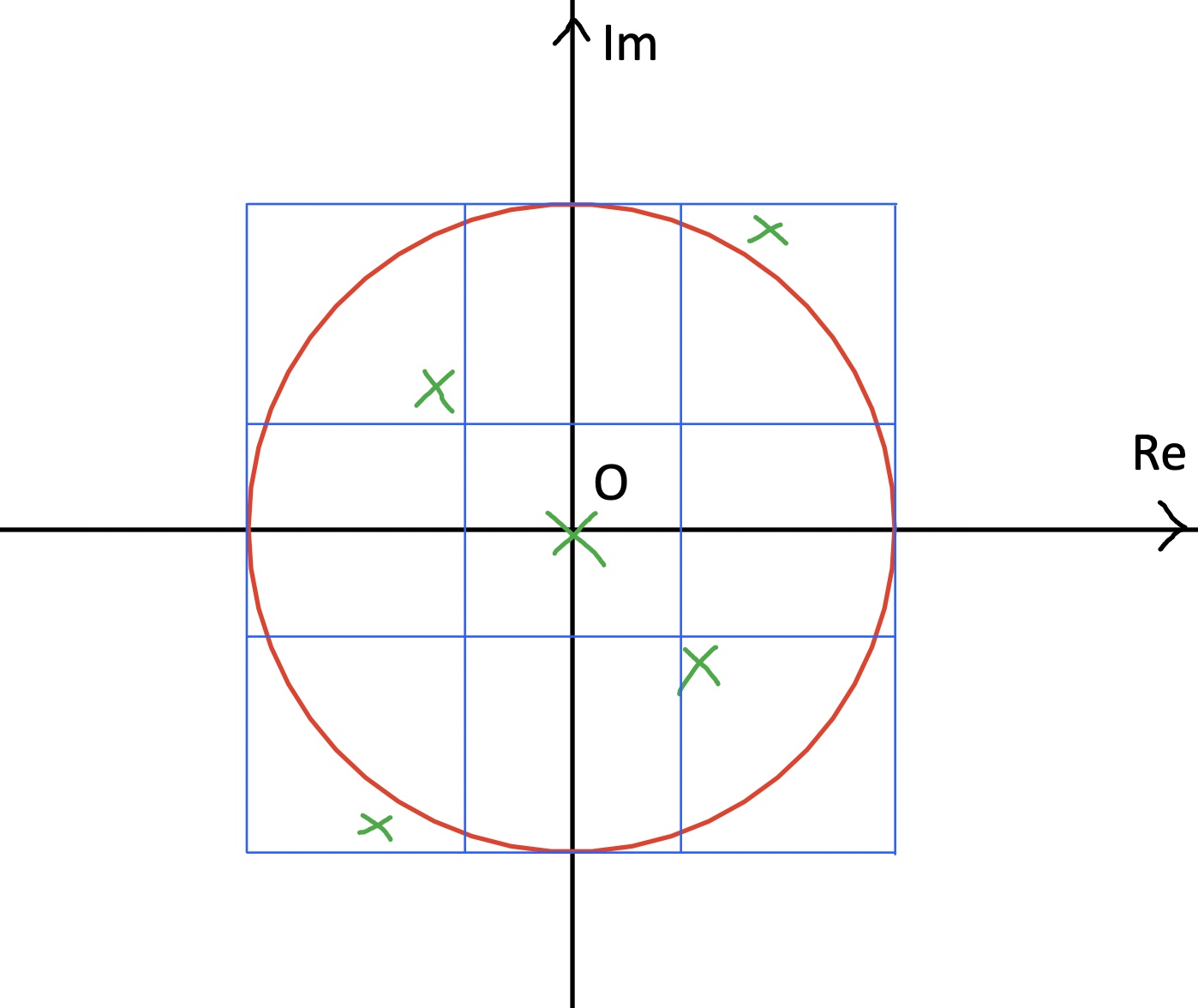}}
    \hfill
    \subcaptionbox{}
    {\includegraphics[scale=0.11]{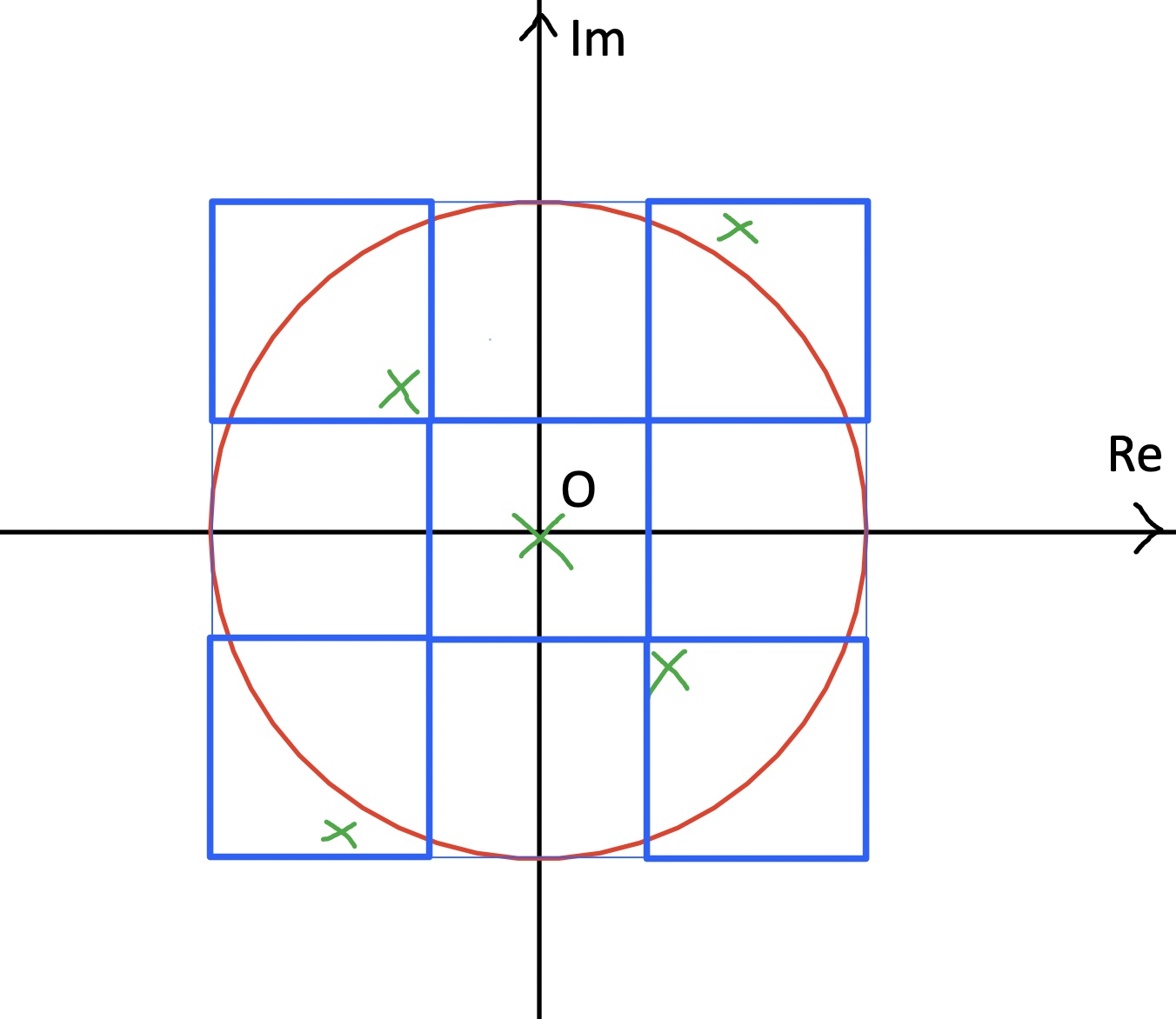}}
    \caption{Illustration of the resolution of the $\xi_n$ search algorithm by dichotomy. The green points correspond to the zeros that the algorithm is looking for. The curves correspond to the integration laces. (a) Finding the initial circle containing the desired number of solutions, (b) breakdown of the domain into 9 squares, (c) only squares containing at least one zero are kept for the next step.}
\label{illustrationDichotomie}
\end{figure}

The algorithm presented allows us to determine the first $\xi_n$, classified by their module. The first step is to determine a domain containing only these $N$ solutions in order to be able to concentrate the searches. Since the aim is to find the solutions with the lowest module, it is natural to look for a domain which is a disc centred on the origin of the complex plan. The initial radius is chosen as a random number between 0 and 1 to avoid the integration yaw passing over a root. The number of solutions in the circle is evaluated numerically by approximating the circle with a 200-sided polygon. As long as the number of roots in the circle is less than the desired number, the radius of the circle is doubled. Once a circle with enough solutions is found, a dichotomy is performed to find a radius with exactly the right number of solutions. This circle is illustrated for $N=3$ in Figure \ref{illustrationDichotomie}(a),and the script \ref{rayon} implemented in Matlab.\\
Once this radius has been determined, the domain is divided into 9 squares. The natural choice of dividing the domain into 4 squares would imply that the integration paths would pass through the point $z=0$. However, at $z=0$, the matrix whose the determinant is evaluated in Equation \eqref{xin} has linearly dependent columns as the third column is equal to the null vector. Thus the determinant is zero and therefore $z=0$ is always a solution of this equation, which would imply that the integration path would pass through a root, preventing the integral of path from being performed.  The squares containing a zero are then kept and the same 9-square division is repeated until the desired accuracy is obtained. The division and the choice of squares to keep are illustrated in Figures \ref{illustrationDichotomie}(b) and \ref{illustrationDichotomie}(c), and the script \ref{solution} implements it in Matlab. The Listing \ref{solutionTriee} then eliminates solutions that are not contained in the initial circle.

The advantage of square laces is that they allow easy meshing, but the convergence of solutions is poor. On the contrary, circular laces converge exponentially, but the meshing of the plan is difficult \cite{CarrePasBien}. An algorithm taking better advantage of circular laces should be considered to obtain more precise results more quickly.

\subsection{Solving the equation for the $A_n$}
 The first step in solving Equation \eqref{An} is to discretise $r/R$ into $N$ regularly spaced points, i.e. $r/R=\lambda/N$, $\lambda=0,1,\dots,(N-1)$. For $\lambda=0$, the last equation is identically zero, so it can be ignored in this case. For $\xi_n=0$, the coefficients of the $A_n$ are zero, so this $\xi_n$ is ignored. The system is now reduced to $3N-1$ equations, which are a priori linearly independent equations. Only $3N-1$ coefficients $A_n$ can therefore be calculated exactly, so it is decided to truncate the series at $3N-1$. The system is now reformulated as a linear equation in the form of Equation \eqref{AnMatrice}.

\begin{equation}
    Ax=b
\label{AnMatrice}
\end{equation}
where $A$ is a square matrix of dimension $(3N-1)\times(3N-1)$, $x$ is a vector containing the $A_n$ and $b$ is the vector containing the constant terms of the system of equations \eqref{An}. The script \ref{coefficient} implements $A$ and $b$ in a Matlab function. Matlab is used for efficient matrix computation, but it is envisaged that the program will be written in Python to facilitate its use in an engineering environment.\\
The system, as it stands, has a conditioning problem. The conditioning (ratio of the largest to the smallest eigenvalue) calculated with Matlab's \verb+cond+ function exceeds $10^{200}$, which is much higher than 1 which is the desired value. To solve this problem, the system is solved for  $A_n\cosh(\xi_nh/R)J_2(\sqrt{\xi_n^2-i\Omega^*})$ instead of $A_n$ in order to regularise these divergent factors in the $A$ matrix. The conditioning is now between $10^{16}$ and $10^{18}$. Since the conditioning is still too high to derive a unique and reliable solution, the method using the least squares with a modified gradient conjugate method (fonction \verb+lsqr+ de Matlab) is used to determine the solution that minimises $||Ax-b||$.

The maximum force according to $x$ and the maximum height can now be calculated numerically.

\section{Results and discussion}
As many approximations have been made, both analytically and numerically, it is necessary to check that the algorithm converges. Figures \ref{convergenceN}(a) and \ref{convergenceN}(b) present a study of convergence in $N$ for $F_{x,max}$ and for $h_{max}$ for demineralised water with $\Omega=10$ $\rm{rad}\cdot s^{-1}$ and $X_0=1$ cm. The maximum $N$ of all the scans by $N$ mentioned in this section is the $N$ at which the algorithm no longer gives physical results (NaNs or $\infty$). The convergence is fast up to about $N=15$, then saturates around a finite value. These values of a few millimetres in height and a few newtons in force are physically plausible results; future experiments will have to be conducted to test these results. From $N=32$, the height decreases towards 0 as $N$ increases. This effect may be due to the fact that the algorithm is trying to invert increasingly large matrices, so it may lose accuracy mainly due to the poor conditioning of the system. The scans in $N$ for the same system with pulses $\Omega=20$ $\rm{rad}\cdot s^{-1}$, $\Omega=23$ $\rm{rad}\cdot s^{-1}$, $\Omega=40$ $\rm{rad}\cdot s^{-1}$, $\Omega=50$ $\rm{rad}\cdot s^{-1}$ et $\Omega=60$ $\rm{rad}\cdot s^{-1}$ are shown in Figures  \ref{SwapN_Omega_20}(a), \ref{SwapN_Omega_20}(b), \ref{SwapN_Omega_30}(a), \ref{SwapN_Omega_30}(b), \ref{SwapN_Omega_40}(a), \ref{SwapN_Omega_40}(b), \ref{SwapN_Omega_50}(a), \ref{SwapN_Omega_50}(b), \ref{SwapN_Omega_60}(a) and \ref{SwapN_Omega_60}(b) shown in Appendix B.\ref{7_1}. As $\Omega$ increases, the algorithm converges less and less quickly and cleanly. From $\Omega=40$ $\rm{rad}\cdot s^{-1}$, the algorithm no longer converges. However, the order of magnitude remains constant for a given impulse. In the following, therefore, $N$ is taken to be equal to 30 if the algorithm gives a result, otherwise it is taken at the last iteration, which gives a finite result.

Figures \ref{SwapEta}(a) and \ref{SwapEta}(b) show that from a dynamic viscosity equal to $0.01$ Pa$\cdot$s (10 times that of demineralised water at 20$^{\rm{o}}$C), $F_{x,max}$ grows linearly with the viscosity. However, the viscosity variations observed are inferior to $1\%$, so the force can be considered as constant according to $\eta$, given the accuracy of the algorithm. The maximum height decreases significantly up to a viscosity of $\eta=0.01$ P$\cdot$s. This attenuation can greatly reduce splashing and non-linear oscillations by ensuring a viscosity above this threshold, which can be evaluated for the desired liquid.

Figure \ref{SwapOmega}(a) shows that the maximum force along $x$ increases with frequency, as expected as additional energy is injected into the system as $\Omega$ becomes large. Around $\Omega=$25 $\rm{rad}\cdot s^{-1}$ there is a small peak on the graph. This local maximum may be due to a resonant frequency that exerts large forces on the walls when excited. Figure \ref{SwapOmega}(b) shows that the maximum height has two significant peaks around $\Omega=$15 $\rm{rad}\cdot s^{-1}$ and $\Omega=$25 $\rm{rad}\cdot s^{-1}$. These peaks may be due to resonant frequencies excitable by this type of excitation depending on $x$. The slight increase towards $\Omega=$65 $\rm{rad}\cdot s^{-1}$ is probably due to the convergence of the algorithm which is no longer perfect in the neighbourhood of these frequencies as discussed earlier.

Figure \ref{SwapAmlp}(a) shows that the maximum force along $x$ increases linearly with the amplitude. The maximum height reaches its maximum $h_{max,max}=1.6$ cm for an amplitude of 3 cm, then decreases significantly to stabilise at  $h_{max}=1$ cm. The excitation amplitude doesn't seem to play a role beyond this threshold.

\section{Conclusion for forced oscillations}
Thus, an algorithm has been designed and numerically implemented to determine the maximum force as a function of $x$ and the maximum height of the free surface of the liquid in a vertical cylindrical tank subjected to a sinusoidal forced horizontal excitation. Convergence studies have been carried out and have concluded that the algorithm converges well for pulsations lower than $30\ \rm{rad}\cdot s^{-1}$. For higher pulsations, the order of magnitude is good, but the algorithm still needs to be improved to achieve good convergence, in particular by preconditioning the system.  A method to determine viscosity thresholds to avoid splashing has been established. Finally, the responses of the liquid for different frequencies and amplitudes of excitation has been established.

\chapter{Conclusion}

Thus, a theoretical model describing the resonant frequencies in rectangular and cylindrical tanks was established. It was verified experimentally in the rectangular case. Algorithms have been implemented to predict the resonant frequencies of the Bella-Lui rocket fuel. The ERT must prevent as much as possible vibrations with a frequency close to these resonant frequencies and ensure that the natural frequency of the rocket does not coincide with one of these resonant frequencies in order to avoid any risk of a Pogo effect.\\
In order to limit the risks to the rocket, the mitigation of liquid sloshing was studied. The previously determined resonant frequencies were used to construct a theoretical model for the free attenuations of viscous fluids. Although the model is not very robust and the results are difficult to analyse quantitatively, qualitative conclusions can be drawn. The higher the resonant frequency, the higher the attenuation associated with that mode. The ERT must therefore reduce the low frequency excitations as much as possible. Adding a layer of a more viscous liquid, even a thin one, will also significantly increase the attenuation coefficient. Moreover, the addition of a layer of polystyrene beads prevents excitation at low frequencies. However, experiments are still needed to find out exactly which physical phenomena explain this.
These two solutions have the advantage of being relatively simple without increasing the mass of the rocket which is adapted to the ERT needs. In addition, the beads are blocked by a filter, preventing them from entering the engine's combustion chamber.\\
Finally, an algorithm was designed and implemented in a Matlab program to determine the behaviour of the liquid when subjected to a horizontal sinusoidal excitation. This is a model that accurately models the response of the liquid when the fuel tank of a rocket is subject to wind gusts or sudden excitation due to the engine or the aerodynamics. The results of the algorithm are promising, but it still encounters convergence problems for extreme cases. Nevertheless, the algorithm can now be used by the ERT to make theoretical predictions for their rockets.

\part*{Appendix}
\appendix

\addcontentsline{toc}{part}{Appendix}

\label{Appendix}

\chapter{Resonant frequencies}

\section{Analytical approach}
\subsection{Physics of fluids}
\label{1_1_1}
The various physical quantities will be expressed in the ground reference frame with, at first, the coordinate system $(x,y,z,t)$ linked to this reference frame.\\
It is assumed that the rocket remains aligned with the  $\hat{\mathbf{z}}$ axis during the entire time the tank is not empty. Its height, velocity and acceleration depend only on time.\\
The liquid propellant contained in the rocket tank is a Newtonian fluid. It is therefore described by the Navier-Stokes Equation given by the Equation  \eqref{NavierStokes} \cite{Meister}.

\begin{equation}
    \rho \textbf{ a} - \nabla P - \rho \nabla \frac{u^2}{2} + \eta {\nabla}^{2} \textbf{u} + (\eta+{\eta}^{*}) \nabla (\nabla \cdot \textbf{u}) = \rho  \textbf{ } \frac{\partial \textbf{u}}{\partial t} - 2 \textbf{ }\rho \textbf{ u } \wedge \textbf{ T}
\label{NavierStokes}
\end{equation}
for which the vortex pseudo-vector $\textbf{T}$ is given by Equation \eqref{Tourbillon}.
\begin{equation}
    \textbf{T}=\frac{1}{2} \nabla \wedge \textbf{u}
\label{Tourbillon}
\end{equation}
The ergol also respects the Continuity Equation given by Equation \eqref{Continuité} \cite{Meister}.
\begin{equation}
    \frac{\partial \rho}{\partial t}+ \nabla \cdot (\rho \textbf{u}) =0
\label{Continuité}
\end{equation}
The only conservative force on the liquid is the earth gravity, positively oriented along the $\hat{\mathbf{z}}$ axis, the acceleration field $\mathbf{a}$ of which is given by Equation \eqref{pesanteur}.
\begin{equation}
    \textbf{a} = g \textbf{ e}_z
    \label{pesanteur}
\end{equation}

To obtain an analytical solution, the ergol is assumed to have the following properties :
\begin{itemize}
    \item perfect ($\eta =\eta^*$ = 0) \cite{Meister}
    \item incompressible ($\chi = 0 \Rightarrow d\rho = \chi \rho dp = 0 \Rightarrow \rho = \mathrm{const}$) \cite{Meister}
    \item irrotational ($\textbf{T} = 0 \Rightarrow  \nabla \wedge \textbf{u} = 0 \Rightarrow \textbf{u} = - \nabla \Phi$) \cite{Meister}
\end{itemize}
The equation \eqref{Continuité} therefore allows us to obtain Equation (1.4)  \eqref{Continuité2},

\begin{equation}
    \nabla \cdot \textbf{u} = 0
    \label{Continuité2}
\end{equation}
and thus Equation \eqref{Continuité3}.
\begin{equation}
    \nabla^2 \Phi = 0 
    \label{Continuité3}
\end{equation}
It is therefore a matter of solving a Laplace equation \cite{Farhat}.
\\
Equation \eqref{NavierStokes} allows us to obtain the Equation   \eqref{NavierStokes2}.
\begin{equation}
    \nabla \left(g z + \frac{P}{\rho} + \frac{1}{2}{(\nabla \Phi \cdot \nabla \Phi)} - \frac{\partial \Phi}{\partial t}\right) = 0
    \label{NavierStokes2}
\end{equation}
By integrating Equation \eqref{NavierStokes2}, Equation \eqref{NavierStokes3} is obtained.
\begin{equation}
    g z + \frac{P}{\rho} + \frac{1}{2}{(\nabla \Phi \cdot \nabla \Phi)} - \frac{\partial \Phi}{\partial t} = C(t)
    \label{NavierStokes3}
\end{equation}
with $C(t)$ a function of times.\\
From now on, the physical quantities will be expressed in the ground reference frame, but with the coordinate system $(x',y',z')$ with as origin the centre of the disc which is the free surface when it is perfectly horizontal..\\
In this new coordinate system, the gradient opertor remains unchanged, but the partial derivative with respect to time in the ground coordinate system is given by Equation \eqref{derivepartieltemps}.
\begin{equation}
    {\left(\frac{\partial }{\partial t}\right)}' = \left(\frac{\partial}{\partial t} - \textbf{V}_0 \cdot \nabla \right)
    \label{derivepartieltemps}
\end{equation}
Equation \eqref{NavierStokes3} then becomes Equation \eqref{NavierStokes4}.
\begin{equation}
    g z + \frac{P}{\rho} + \frac{1}{2}{(\nabla \Phi \cdot \nabla \Phi)} - \frac{\partial \Phi}{\partial t} + \textbf{V}_0 \cdot \nabla\Phi  = C(t)
    \label{NavierStokes4}
\end{equation}
At the surface of the ergol ($z'=\xi$), the pressure is given by Equation \eqref{pression} \cite{Ibrahim}.
\begin{equation}
    P = p_{ext} + p_{s}
    \label{pression}
\end{equation}
with the Laplace pressure $p_s$ given by the Laplace-Young equation : $p_s = - \gamma \kappa = \gamma (\frac{1}{R_1} + \frac{1}{R_2})$.\\
Considering Equation \eqref{changementCoordonnees}, Equation \eqref{NavierStokes4} becomes Equation \eqref{NavierStokes5}.
\begin{equation}
    g \xi + \frac{\gamma }{\rho}\left(\frac{1}{R_1} + \frac{1}{R_2}\right) + \frac{1}{2}{(\nabla \Phi \cdot \nabla \Phi)} - \frac{\partial \Phi}{\partial t} + \textbf{V}_0 \cdot \nabla\Phi  = C'(t)
    \label{NavierStokes5}
\end{equation}
for which $C'(t) = C(t)- \frac{P}{\rho} - \textbf{V}_0 t$ is a function of time.\\
By gauge invariance (changing $\Phi$ en $\left[\Phi+\int C'(t)dt\right]$ does not change the gradient of the potential, the physics of the system are therefore not impacted by this change), Equation, \eqref{NavierStokes5} is equivalent to Equation \eqref{NavierStokes6}.
\begin{equation}
    g \xi + \frac{\gamma }{\rho}\left(\frac{1}{R_1} + \frac{1}{R_2}\right) + \frac{1}{2}{(\nabla \Phi \cdot \nabla \Phi)} - \frac{\partial \Phi}{\partial t} + \textbf{V}_0 \cdot \nabla\Phi  = 0
    \label{NavierStokes6}
\end{equation}
Equation \eqref{phiTildeLaplace} shows that $\Tilde{\Phi}$ is governed by Laplace's equation.
\begin{equation}
\begin{split}
    \nabla^2\Tilde{\Phi}&=\nabla^2(\Phi-\Phi_0)\\
    &=0-\nabla\cdot (-\nabla\Phi_0)\\
    &=\nabla\cdot {V_0}\\
    &=0
\label{phiTildeLaplace}
\end{split}
\end{equation}
Equation \eqref{potentiel} separates the potential into a disturbance potential $\Tilde{\Phi}$ and a potential linked to the movement of the tank  $\Phi_0$, such that $\textbf{V}_{0} = -\nabla \Phi_0 $.
\begin{equation}
    \Phi = \Tilde{\Phi} + \Phi_0
    \label{potentiel}
\end{equation}
Equation \eqref{NavierStokes6} thus becomes the Equation \eqref{NavierStokes7}.
\begin{equation}
    g \xi + \frac{\gamma }{\rho}\left(\frac{1}{R_1} + \frac{1}{R_2}\right) + \frac{1}{2}{(\nabla \Tilde{\Phi} \cdot \nabla \Tilde{\Phi})} - \frac{\partial \Tilde{\Phi}}{\partial t} - \frac{\partial \Phi_0}{\partial t} - \frac{1}{2} {V_0}^2= 0
    \label{NavierStokes7}
\end{equation}

\subsection{Rectangular tank}
\label{1_1_2}
Let be a rectangular tank of length $L$, width $l$ and such that the height of the fluid at rest is  $H$ .\\
Placing ourselves in a Cartesian coordinate system with as origin the centre of the rectangle which is the free surface when it is perfectly horizontal, $V_0$ is expressed by the Equation \eqref{vitesseR}.
\begin{equation}
    V_0 =\Dot{Z_0} \textbf{e}_{z}
    \label{vitesseR}
\end{equation}
Considering the boundary conditions given by Equation  \eqref{bord1R}, Equation \eqref{vitesseR} is integrated with respect to the spatial coordinates to obtain Equation \eqref{potentiel0R}.
\begin{equation}
    -\left.\frac{\partial \Phi_0}{\partial z}\right|_{z=-H} = \Dot{Z_0}
    \label{bord1R}
\end{equation}
\begin{equation}
    \Phi_0 = \Dot{Z_0}z
    \label{potentiel0R}
\end{equation}
By explicitly expressing the curvature in these coordinates, the Equation \eqref{CourbureR} is obtained \cite{Ibrahim}.
\begin{equation}
\begin{split}
    - \kappa &= \left(\frac{1}{R_1} + \frac{1}{R_2}\right) \\  
    &= \frac{\xi_{xx}\left(1 + {\xi_{y}}^{2}\right) + \xi_{yy}\left(1 + {\xi_{x}}^{2}\right) - 2 \xi_{x} \xi_{y} \xi_{xy}}{{\left(1 + {\xi_{y}}^{2} + {\xi_{x}}^{2}\right)}^{\frac{3}{2}}}
    \label{CourbureR}
\end{split}
\end{equation}
Equation \eqref{CourbureR} can be linearised to obtain Equation \eqref{CourbureR1}.
\begin{equation}
    - \kappa = (\xi_{xx} + \xi_{yy})
    \label{CourbureR1}
\end{equation}
The Equation \eqref{NavierStokes7} thus becomes Equation \eqref{NavierStokesR1}.
\begin{equation}
    (g + \Ddot{Z_0}) z + \frac{\gamma }{\rho}\left(\frac{{\partial}^2 z}{{\partial x}^2} + \frac{{\partial}^2 z}{{\partial y}^2}\right) + \frac{1}{2}{(\nabla \Tilde{\Phi} \cdot \nabla \Tilde{\Phi})} - \frac{\partial \Tilde{\Phi}}{\partial t}
    = 0 \textrm{, à }z = \xi(x,y,t)
    \label{NavierStokesR1}
\end{equation}

At the free surface, the velocity of a fluid particle is equal to the vertical velocity of the free surface. This condition is the Equation \eqref{interfaceLibreR}, known as the kinematic condition of the free surface.
\begin{equation}
    -\frac{\partial \Phi}{\partial z}=\frac{\partial z}{\partial t}+\textbf{u}_{rel}\cdot\nabla z \textrm{, à }z = \xi(x,y,t).
\label{interfaceLibreR}
\end{equation}
The Equation \eqref{derivephizR} is obtained by considering the gradient in Cartesian coordinates and $\textbf{u}_{rel}=-\nabla \ \Tilde{\Phi}$.
\begin{equation}
    -\frac{\partial \Tilde{\Phi}}{\partial z} =  \frac{\partial z}{\partial t} - \frac{\partial z}{\partial x}\frac{\partial \Tilde{\Phi}}{\partial x} - \frac{1}{r^2} \frac{\partial z}{\partial y}\frac{\partial \Tilde{\Phi}}{\partial y}\textrm{, à }z = \xi(x,y,t).
    \label{derivephizR}
\end{equation}
By variational approach, using the framework of Laplace's equation and using the boundary conditions given by the Equations \eqref{bord2R},\eqref{bord3R},\eqref{bord4R}, a solution for $ \Tilde {\Phi}$ is obtained, the form of which is given by Equation \eqref{solutionR}.
\begin{equation}
    -\left.\frac{\partial \Tilde{\Phi}}{\partial z}\right|_{z=-H} = 0
    \label{bord2R}
\end{equation}
\begin{equation}
    -\left.\frac{\partial \Tilde{\Phi}}{\partial x}\right|_{z=\pm \frac{L}{2}} = 0
    \label{bord3R}
\end{equation}
\begin{equation}
    -\left.\frac{\partial \Tilde{\Phi}}{\partial y}\right|_{z=\pm \frac{l}{2}} = 0
    \label{bord4R}
\end{equation}
\begin{equation}
    \Tilde{\Phi}(r,\theta,z,t) = \sum_{m=0}^{\infty} \sum_{n=1}^{\infty} \left[\Bar{\alpha}_{mm}(t) \cos\left(\frac{2 m  \pi x}{L}\right) \cos\left(\frac{2 m  \pi y}{l}\right)\right] \cosh(k_{mn}(z + h))
    \label{solutionR}
\end{equation}
where $k_{mn} = \pi \sqrt{((2m)^2/L^2) + (2n)^2/l^2)}$.\\
From now on small oscillations will be considered. The vertical acceleration of the rocket and the height of the liquid in the tank are assumed to vary slowly enough to be considered as constants.\\
By making these approximations, Equation \eqref{NavierStokesR1} can be reduced to Equation \eqref{NavierStokesR2}.
\begin{equation}
    (g + \Ddot{Z_0})\frac{{\partial} \Phi}{{\partial z}} + \frac{\gamma}{\rho}\frac{{\partial}^{2} \Phi}{{\partial t}^{2}} + \frac{{\partial}^{2} \Phi}{{\partial t}^{2}} = 0 
    \label{NavierStokesR2}
\end{equation}
Considering that the functions $\alpha_{mn}(t)$ are expressed as harmonics of the form  $\sin(\omega_{mn}t)$ and after having used Laplace's equations, the eigen-pulsations $\omega_{mn}$ given by Equation  \eqref{pulsationR} are obtained.
\begin{equation}
    \omega_{mn}^{2} = \left[(g + \Ddot{Z_0}) k_{mn} + \frac{\gamma}{\rho} {k_{mn}}^{3} \right] \tanh\left(k_{mn} h\right)
    \label{pulsationR}
\end{equation}
The eigen-frequencies are therefore given by the Equation \eqref{fréquenceR}.
\begin{equation}
    f_{mn} = \frac{1}{2 \pi}\sqrt{\left[(g + \Ddot{Z_0}) k_{mn} + \frac{\gamma}{\rho} {k_{mn}}^{3} \right] \tanh\left(k_{mn} h\right)}
    \label{fréquenceR}
\end{equation}

\subsection{Cylindrical tank}
\label{1_1_3}
Let be a cylindrical tank of radius $R$ and such that the height of the fluid at rest is $h$.\\
Placing ourselves in a cylindrical coordinate system with as origin the centre of the disc which is the free surface when it is perfectly horizontal, $V_0$ is expressed by the Equation \eqref{vitesseR}.
\begin{equation}
    V_0 =\Dot{Z_0} \textbf{e}_{z}
    \label{vitesseC}
\end{equation}
Considering the boundary conditions given by Equation \eqref{bord1C}, Equation \eqref{vitesseC} is integrated with respect to the spatial coordinates to obtain Equation \eqref{potentiel0C}.
\begin{equation}
    -\left.\frac{\partial \Phi_0}{\partial z}\right|_{z=-h} = \Dot{Z_0}
    \label{bord1C}
\end{equation}
\begin{equation}
    \Phi_0 = \Dot{Z_0}z
    \label{potentiel0C}
\end{equation}
By expressing the curve explicitly in these coordinates, Equation  \eqref{CourbureC} is obtained \cite{Ibrahim}.
\begin{equation}
\begin{split}
    - \kappa &= (\frac{1}{R_1} + \frac{1}{R_2}) \\  
    &= \frac{\xi_{rr}(1 + ({\xi_{\theta}}^{2}/r^{2})) + (1 + {\xi_{r}}^{2}) (({\xi_{r}}^{2}/r) + ({\xi_{\theta \theta}}/r^{2})) - 2 \xi_{r}({\xi_{\theta}}/r^{2})(\xi_{rr} + (\xi_{\theta}/r))}{{\left(1 + {\xi_{r}}^{2} + ({\xi_{\theta}}^{2}/r^{2})\right)}^{\frac{3}{2}}}
    \label{CourbureC}
\end{split}
\end{equation}
The Equation \eqref{CourbureC} can be linearized to obtain the Equation  \eqref{CourbureC1}.
\begin{equation}
    - \kappa = (\xi_{rr} + {\xi_{r}}/r + {\xi_{\theta \theta}}/r^{2})
    \label{CourbureC1}
\end{equation}
the Equation \eqref{NavierStokes7} thus becomes the Equation \eqref{NavierStokesC1}.
\begin{equation}
    (g + \Ddot{Z_0}) z + \frac{\gamma }{\rho}\left(\frac{{\partial}^2 z}{{\partial r}^2} + \frac{1}{r} \frac{\partial z}{\partial r} + \frac{1}{r^2}\frac{{\partial}^2 z}{{\partial \theta}^2}\right) + \frac{1}{2}{(\nabla \Tilde{\Phi} \cdot \nabla \Tilde{\Phi})} - \frac{\partial \Tilde{\Phi}}{\partial t} 
    = 0 \textrm{, à }z = \xi(r,\theta,t).
    \label{NavierStokesC1}
\end{equation}

At the free surface, the velocity of a fluid particle is equal to the vertical velocity of the free surface. This condition is the Equation \eqref{interfaceLibreC}, known as the kinematic condition of the free surface.
\begin{equation}
    -\frac{\partial \Phi}{\partial z}=\frac{\partial z}{\partial t}+\textbf{u}_{rel}\cdot\nabla z \textrm{, à }z = \xi(r,\theta,t).
\label{interfaceLibreC}
\end{equation}
The Equation \eqref{derivephizC} is obtained by considering the gradient in cylindrical coordinates and $\textbf{u}_{rel}=-\nabla \ \Tilde{\Phi}$.
\begin{equation}
    -\frac{\partial \Tilde{\Phi}}{\partial z} =  \frac{\partial z}{\partial t} - \frac{\partial z}{\partial r}\frac{\partial \Tilde{\Phi}}{\partial r} - \frac{1}{r^2} \frac{\partial z}{\partial \theta}\frac{\partial \Tilde{\Phi}}{\partial \theta}\textrm{, à }z = \xi(r,\theta,t).
    \label{derivephizC}
\end{equation}

By variational approach, using as fact the framework of Laplace equations and using the boundary conditions given by Equations \eqref{bord2C} and \eqref{bord3C}, a solution for $ \Tilde {\Phi}$ whose form is given by Equation \eqref{solutionC} is obtained.
\begin{equation}
    \left.\frac{\partial \Tilde{\Phi}}{\partial z}\right|_{z=-h} = 0
    \label{bord2C}
\end{equation}
\begin{equation}
    \left.\frac{\partial \Tilde{\Phi}}{\partial r}\right|_{z=R} = 0
    \label{bord3C}
\end{equation}

\begin{equation}
    \Tilde{\Phi}(r,\theta,z,t) = \sum_{m=0}^{\infty} \sum_{n=1}^{\infty} [\alpha_{mn}(t) \cos(m \theta) + \beta_{mn}(t) \sin(m \theta)] J_m(\lambda_{mn}r) \frac{\cosh(\lambda_{mn}(z + h))}{\cosh(\lambda_{mn}h)}
    \label{solutionC}
\end{equation}
with $\lambda_{mn} = \varepsilon_{mn} / R$.\\
From now on, small oscillations will be considered. The vertical acceleration of the rocket and the height of the liquid in the tank are assumed to vary sufficiently slowly to be considered as constants.\\
By making these approximations, Equation \eqref{NavierStokesC1} is reduced to Equation \eqref{NavierStokesC2}.
\begin{equation}
    (g + \Ddot{Z_0})\frac{{\partial} \Phi}{{\partial z}} + \frac{\gamma}{\rho}\frac{{\partial}^{2} \Phi}{{\partial t}^{2}} + \frac{{\partial}^{2} \Phi}{{\partial t}^{2}} = 0 
    \label{NavierStokesC2}
\end{equation}
Considering that the functions $\alpha_{mn}(t)$ and $\beta_{mn}(t)$ are expressed as harmonics of the form $\sin(\omega_{mn}t)$ and after using Laplace's equations, the eigen-pulsations $\omega_{mn}$ given by Equation  \eqref{pulsationC} are obtained.
\begin{equation}
    \omega_{mn}^{2} = \left[(g + \Ddot{Z_0})\lambda_{mn} + \frac{\gamma}{\rho } {\lambda_{mn}}^{3} \right] \tanh\left(h {\lambda_{mn}}\right)
    \label{pulsationC}
\end{equation}
The proper frequencies are therefore given by the Equation \eqref{fréquenceC}.
\begin{equation}
    f_{mn} = \frac{1}{2 \pi}\sqrt{\left[(g + \Ddot{Z_0})\lambda_{mn} + \frac{\gamma}{\rho } {\lambda_{mn}}^{3} \right] \tanh\left(h {\lambda_{mn}}\right)}
    \label{fréquenceC}
\end{equation}
\subsection{Differential operator and coordinate change}
Let the coordinate system $(x,y,z,t)$ and the coordinate system $(x',y',z',t)$ be linked by Equation \eqref{changementCoordonnees}.

\begin{equation}
\begin{split}
    x' &= x + V_{0x} t\\
    y' &= y + V_{0y} t\\
    z' &= z + V_{0z} t\\
\end{split}
\label{changementCoordonnees}
\end{equation}
Given a function $f(x,y,z,t)$, its differential is expressed by the Equation \eqref{différentiel} in the reference frame  $(x,y,z,t)$.

\begin{equation}
    \dd f=\frac{\partial f(x,y,z,t)}{\partial x}\dd x+\frac{\partial f(x,y,z,t)}{\partial y}\dd y+\frac{\partial f(x,y,z,t)}{\partial z}\dd z+\frac{\partial f(x,y,z,t)}{\partial t}\dd t\\
\label{différentiel}
\end{equation}
Using Equation \eqref{changementCoordonnees}, the differential of the function $f$ can be expressed by Equation \eqref{différentiel2} in the coordinates   $(x',y',z',t)$.
\begin{equation}
\begin{split}
   \dd f =&\frac{\partial f(x,y,z,t)}{\partial x}(\dd x'-V_{0x}\dd t)+\frac{\partial f(x,y,z,t)}{\partial y}(\dd y'-V_{0y}\dd t)\\
    &+\frac{\partial f(x,y,z,t)}{\partial z}(\dd z'-V_{0z}\dd t)+\frac{\partial f(x,y,z,t)}{\partial t}\dd t\\
    =&\frac{\partial f(x,y,z,t)}{\partial x}\dd x'+\frac{\partial f(x,y,z,t)}{\partial y}\dd y'+\frac{\partial f(x,y,z,t)}{\partial z}\dd z'\\
    &+\left(\frac{\partial f(x,y,z,t)}{\partial t}-V_{0x}\frac{\partial f(x,y,z,t)}{\partial x}-V_{0y}\frac{\partial f(x,y,z,t)}{\partial y}-V_{0z}\frac{\partial f(x,y,z,t)}{\partial z}\right)\dd t
\end{split}
\label{différentiel2}
\end{equation}
The gradient and the partial time derivative in the new coordinates are thus obtained by identification, and are respectively given by Equations \eqref{gradient} and \eqref{derpartemp}.
\begin{equation}
    \nabla f(x',y',z',t)=\nabla f(x,y,z,t)\\
\label{gradient}
\end{equation}
\begin{equation}
    \left(\frac{\partial}{\partial t}\right)f(x',y',z',t)=\left(\frac{\partial}{\partial t}-\textbf{V}_0\cdot\nabla \right)f(x,y,z,t)
\label{derpartemp}
\end{equation}

\section{Experimental Approach}
\subsection{Liquid temperature}
\label{1_2_1}
\begin{figure}[H]
    \centering
    \includegraphics[scale=1]{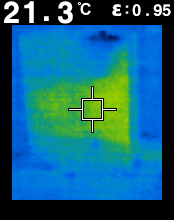}
    \caption{Photo of the tank containing demineralised water taken during the experiment, to which a coloured temperature gradient is applied..}
    \label{reservoir_température}
\end{figure}

The temperature measurement is carried out with a FLIR TG167 camera. The measurement is shown in Figure \ref{reservoir_température}

\subsection{Frequency swaps and other results}
\label{1_2_2}
\begin{figure}[H]
    \centering
    \includegraphics[scale=0.9]{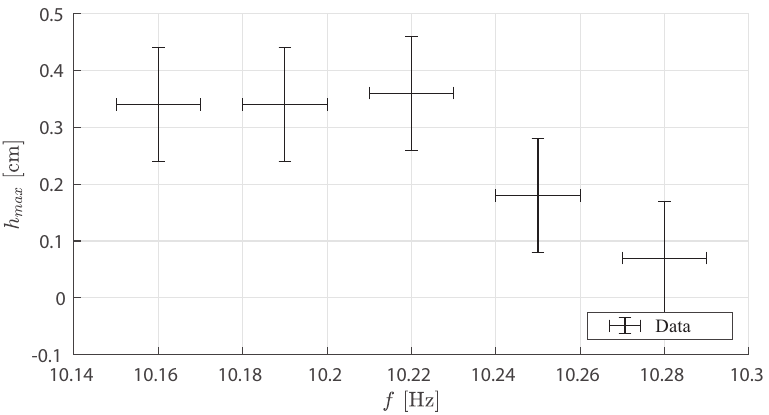}
    \caption{Frequency swap for the $(3,0)$ mode at a height of $h = 1.0$ cm.}
    \label{water_30_h_1_0}
\end{figure}

\begin{figure}[H]
    \centering
    \includegraphics[scale=0.9]{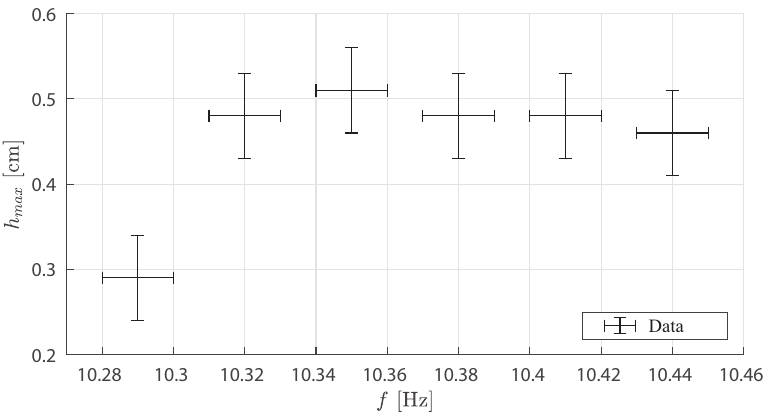}
    \caption{Frequency swap for the $(3,0)$ mode at a height of $h = 1.5$ cm.}
    \label{water_30_h_1_5}
\end{figure}

\begin{figure}[H]
    \centering
    \includegraphics[scale=0.9]{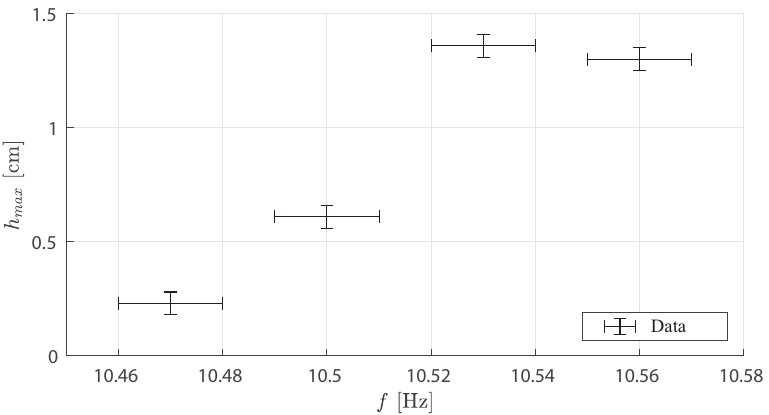}
    \caption{Frequency swap for the $(3,0)$ mode at a height of $h = 2.5$ cm.}
    \label{water_30_h_2_5}
\end{figure}

\begin{figure}[H]
        \centering
        \includegraphics[scale=0.55]{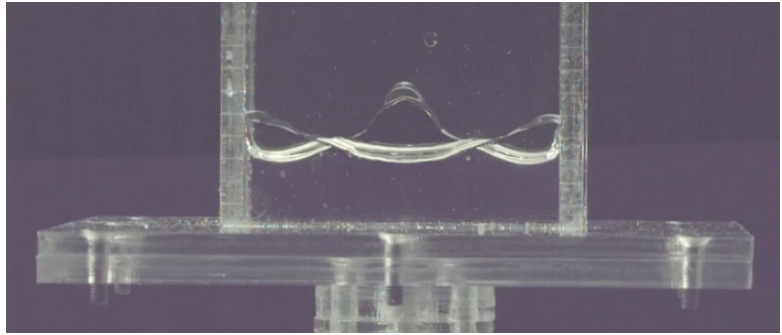}
        \caption{Observation of the $(3,0)$ mode at a frequency of $10.45$ Hz and a height of $2$ cm.}
        \label{f_10_45_h_2_0}
    \end{figure}

The frequency swap results for heights of $h$ de $1.0$ cm, $1.5$ cm et $2.5$ cm are shown in Figures \ref{water_30_h_1_0}, \ref{water_30_h_1_5} and \ref{water_30_h_2_5}.
\paragraph{Frequency swap}
Figure \ref{f_10_45_h_2_0} is the superposition of a photo when the waves are at their lowest and a photo when the waves are at their highest. 
    
\section{Matlab script for cylindrical tank}
\label{1_3_1}
In order to solve numerically the Equation \eqref{fréquenceCA}, it is necessary to implement a function to find the n-th root of the derivative of the m-th Bessel function of the first kind $J_m$. They are pre-calculated for the first $(n,m)$.

\matlabscript{RootBesselFonctionFirstKind}{This script returns the n-th root of the m-th Bessel function of the first kind $J_m$.}
Two functions are then used to evaluate the Equation \eqref{fréquenceCA} at different times during the rocket's flight.

\matlabscript{FrequenceResonance}{This script calculates the resonant frequencies for a given frequency during flight.}

\matlabscript{FrequenceResonanceVol}{This script uses the \ref{FrequenceResonance} script to calculate the first resonant frequencies during flight.}

\chapter{\ Numerical simulations of forced oscillations}
\section{Script Matlab}
\matlabscript{rayon}{Matlab program that determines the radius $r$ in the search algorithm for $\xi_n$.}

\matlabscript{solution}{Matlab program that returns all $\xi_n$ in a square of side $2r$.}

\matlabscript{solutionTriee}{Matlab program that sorts the solutions found.}
    
\matlabscript{coefficient}{Matlab program that creates the matrix $A$ and the vector $b$ such that the $A_n$ are the solution $x$ of the Equation \eqref{AnMatrice}.}

\matlabscript{forceX}{Matlab program that calculates the maximum force according to $x$ and the maximum height of the liquid in forced oscillation.}

\section{Graphics}
\label{7_1}
\begin{figure}[H]
    \centering
    \subcaptionbox{}
    {\includegraphics[scale=0.4]{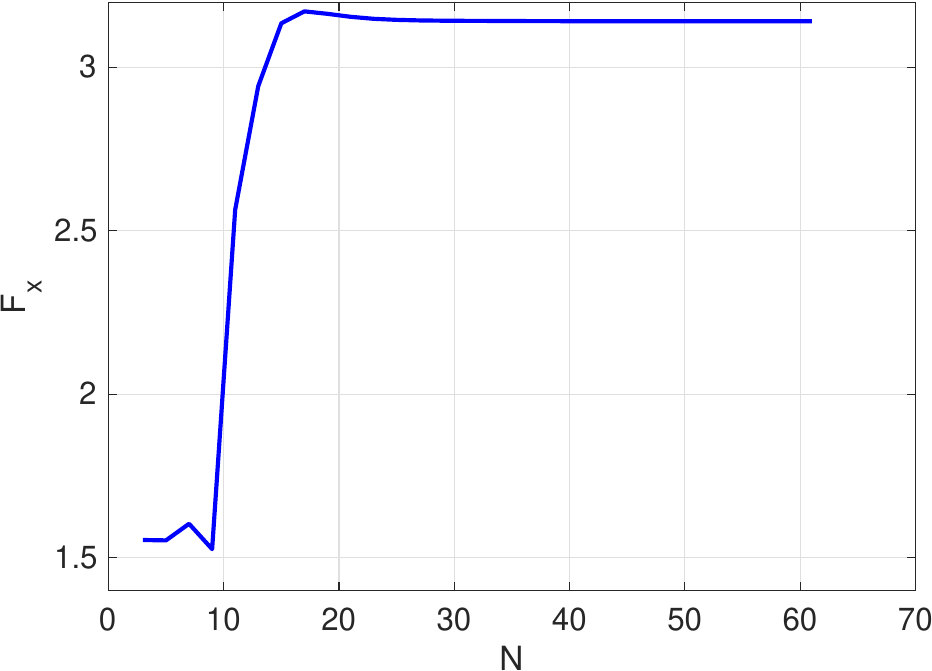}}
    \hfill
    \subcaptionbox{}
    {\includegraphics[scale=0.4]{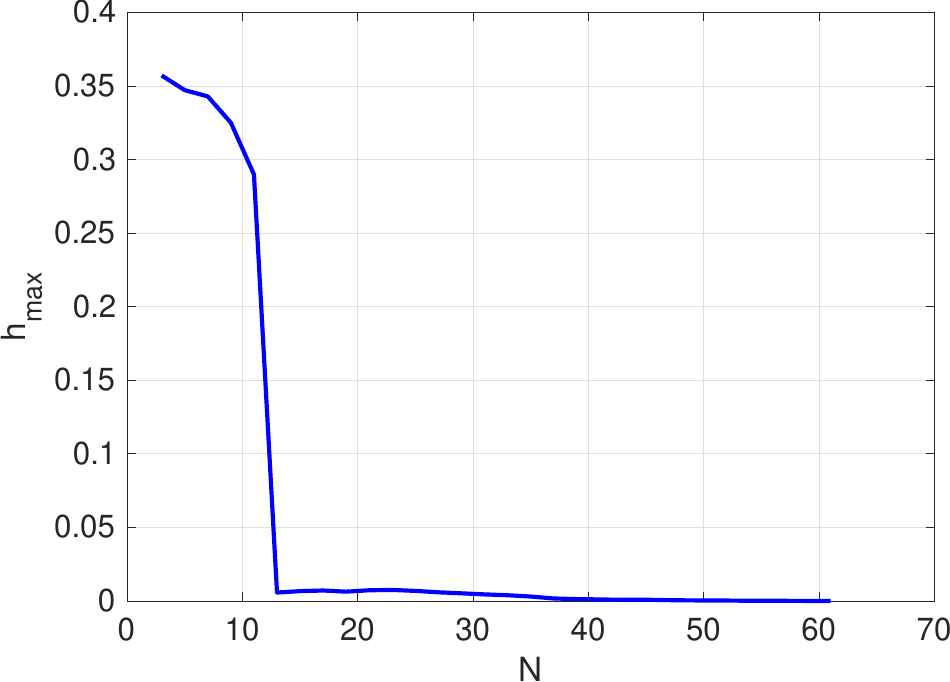}}
    \caption{Convergence study of the algorithm as a function of $N$ for demineralised water with $\Omega=10$ $\rm{rad}\cdot s^{-1}$ et $X_0=1$ cm (a) for $F_{x,max}$ (b) for $h_{max}$.}
\label{convergenceN}
\end{figure}

\begin{figure}[H]
    \centering
    \subcaptionbox{}
    {\includegraphics[scale=0.4]{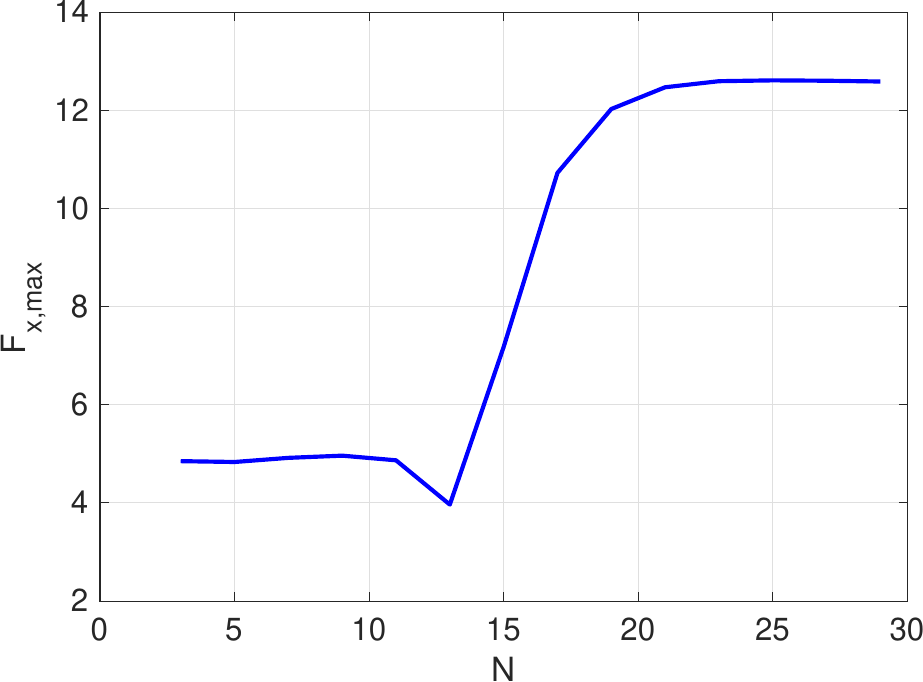}}
    \hfill
    \subcaptionbox{}
    {\includegraphics[scale=0.4]{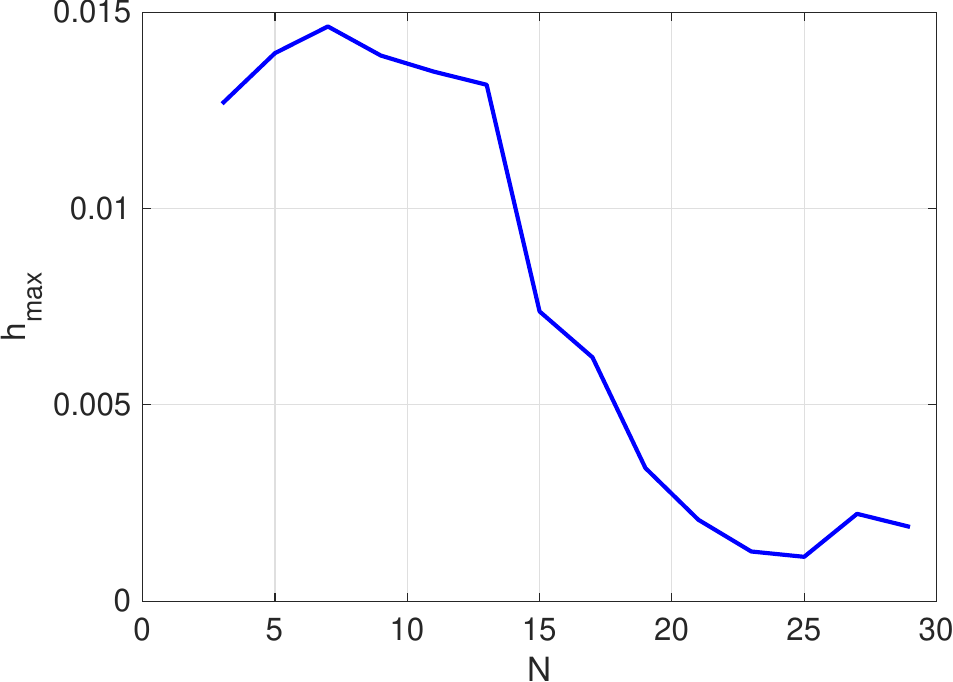}}
    \caption{Convergence study of the algorithm as a function of  $N$ for distilled water with  $\Omega=20$ $\rm{rad}\cdot s^{-1}$ et $X_0=1$ cm (a) for $F_{x,max}$ (b) for $h_{max}$.}
\label{SwapN_Omega_20}
\end{figure}

\begin{figure}[H]
    \centering
    \subcaptionbox{}
    {\includegraphics[scale=0.4]{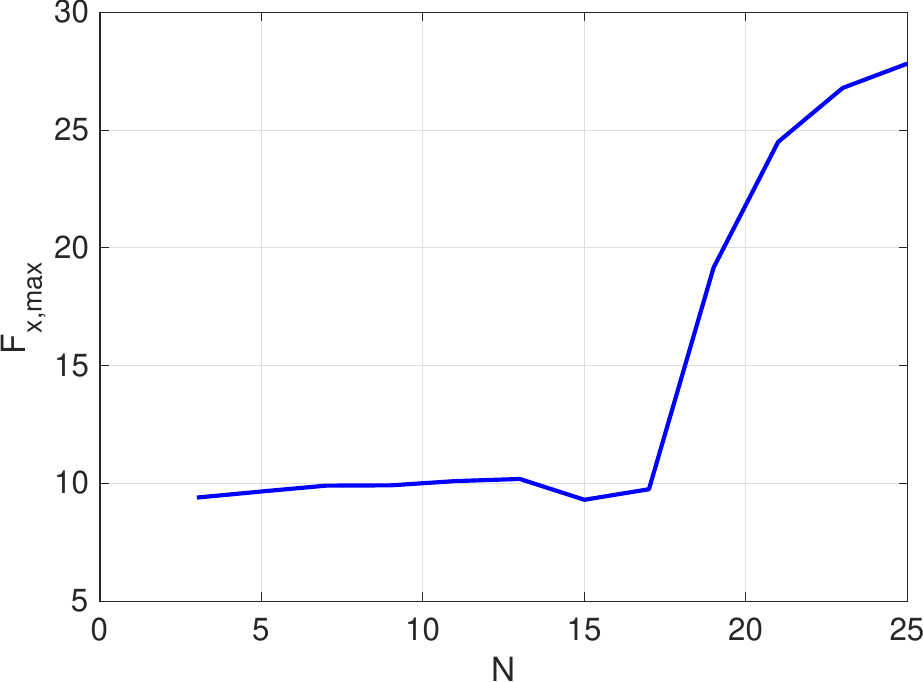}}
    \hfill
    \subcaptionbox{}
    {\includegraphics[scale=0.4]{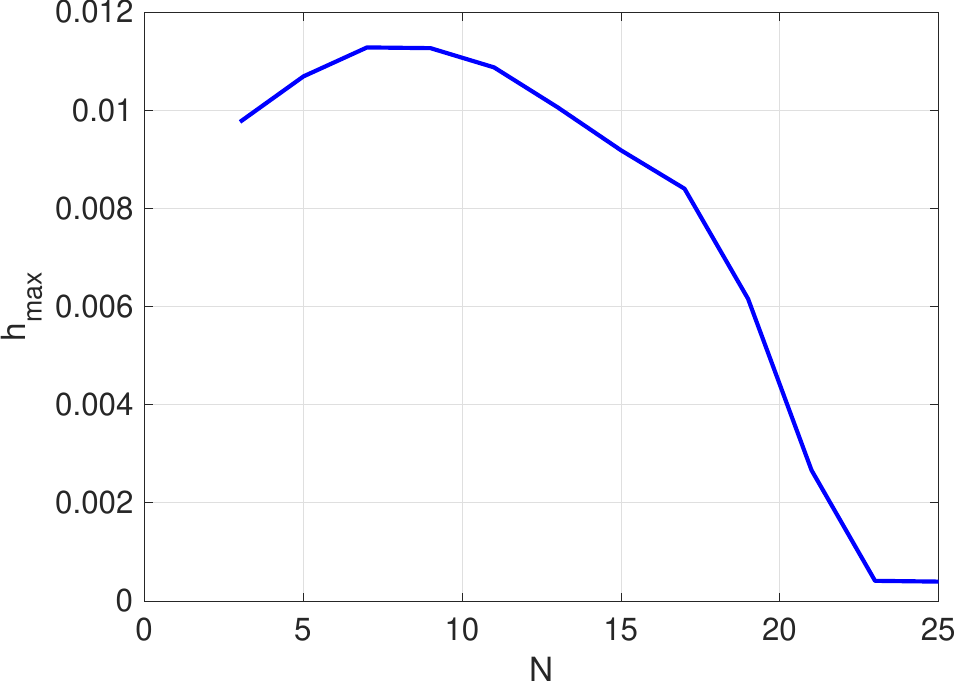}}
    \caption{Convergence study of the algorithm as a function of $N$ for distilled water with  $\Omega=30$ $\rm{rad}\cdot s^{-1}$ et $X_0=1$ cm (a) for $F_{x,max}$ (b) for $h_{max}$.}
\label{SwapN_Omega_30}
\end{figure}

\begin{figure}[H]
    \centering
    \subcaptionbox{}
    {\includegraphics[scale=0.4]{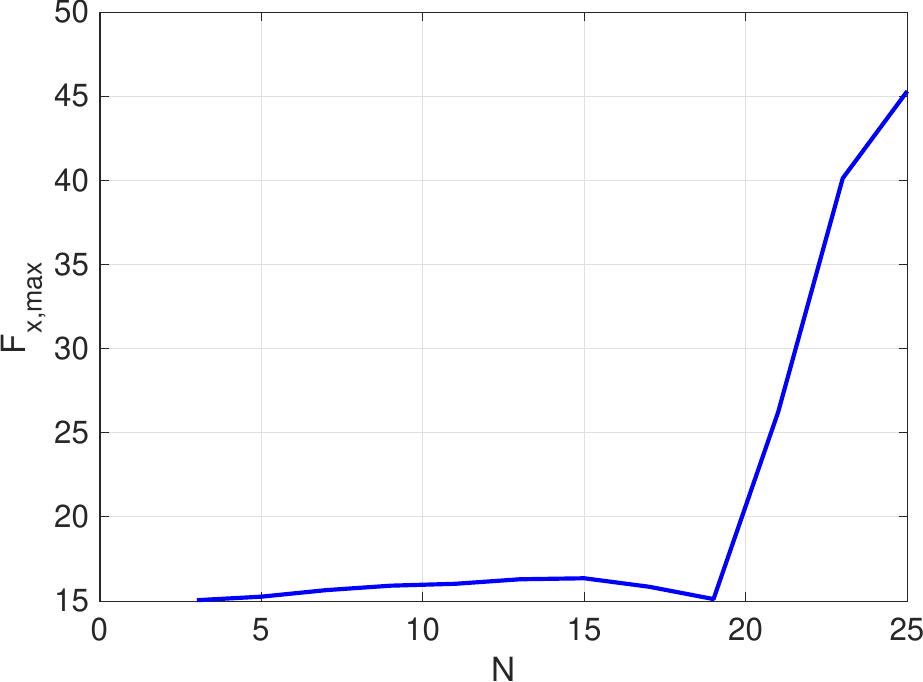}}
    \hfill
    \subcaptionbox{}
    {\includegraphics[scale=0.4]{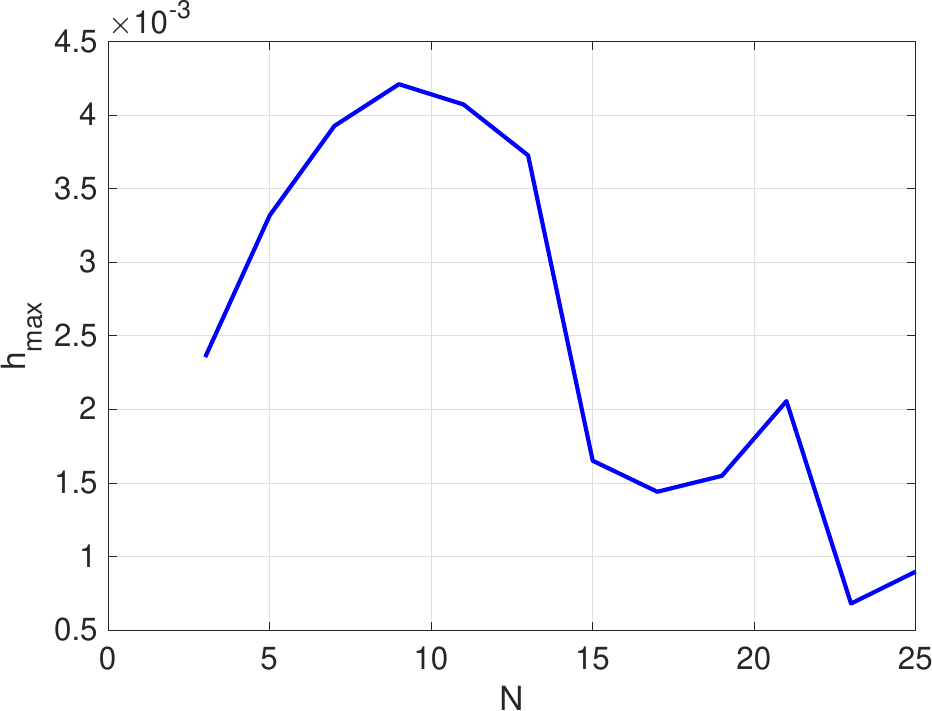}}
    \caption{Convergence study of the algorithm as a function of  $N$ for distilled water with $\Omega=40$ $\rm{rad}\cdot s^{-1}$ et $X_0=1$ cm (a) for $F_{x,max}$ (b) for $h_{max}$.}
\label{SwapN_Omega_40}
\end{figure}

\begin{figure}[H]
    \centering
    \subcaptionbox{}
    {\includegraphics[scale=0.4]{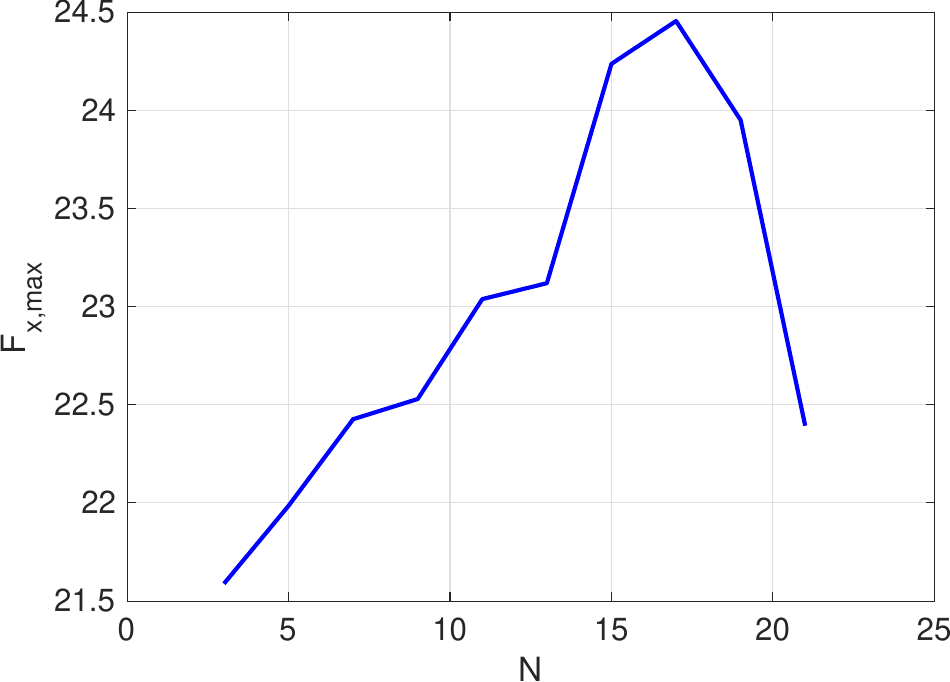}}
    \hfill
    \subcaptionbox{}
    {\includegraphics[scale=0.4]{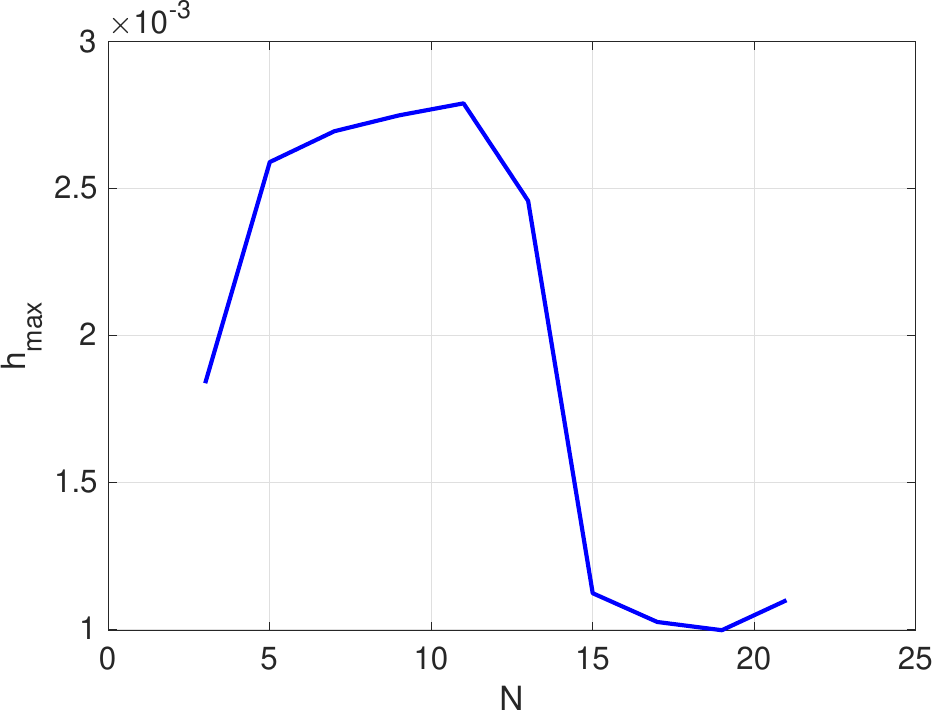}}
    \caption{Convergence study of the algorithm as a function of  $N$ for distilled water with $\Omega=50$ $\rm{rad}\cdot s^{-1}$ et $X_0=1$ cm (a) for $F_{x,max}$ (b) for $h_{max}$.}
\label{SwapN_Omega_50}
\end{figure}

\begin{figure}[H]
    \centering
    \subcaptionbox{}
    {\includegraphics[scale=0.4]{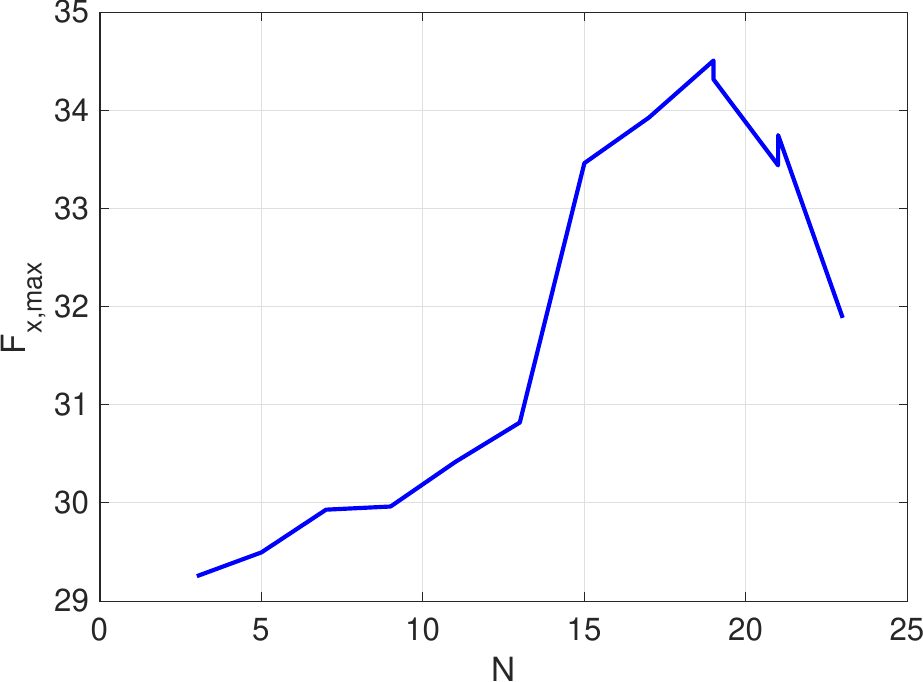}}
    \hfill
    \subcaptionbox{}
    {\includegraphics[scale=0.4]{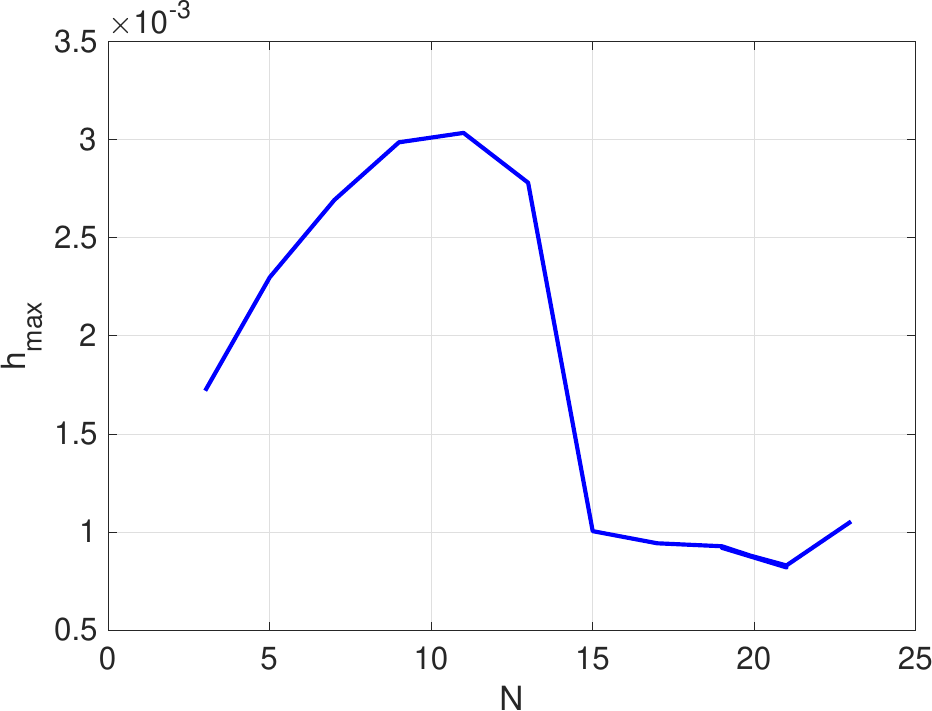}}
    \caption{Convergence study of the algorithm as a function of $N$ for distilled water with $\Omega=60$ $\rm{rad}\cdot s^{-1}$ et $X_0=1$ cm (a) for $F_{x,max}$ (b) for $h_{max}$.}
\label{SwapN_Omega_60}
\end{figure}

\begin{figure}[H]
    \centering
    \subcaptionbox{}
    {\includegraphics[scale=0.4]{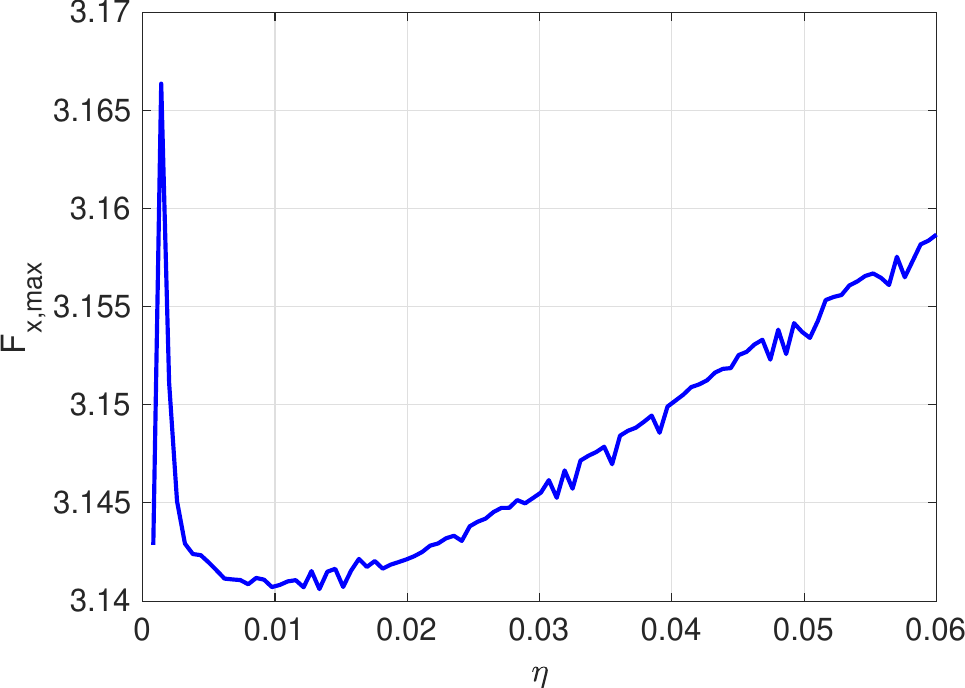}}
    \hfill
    \subcaptionbox{}
    {\includegraphics[scale=0.4]{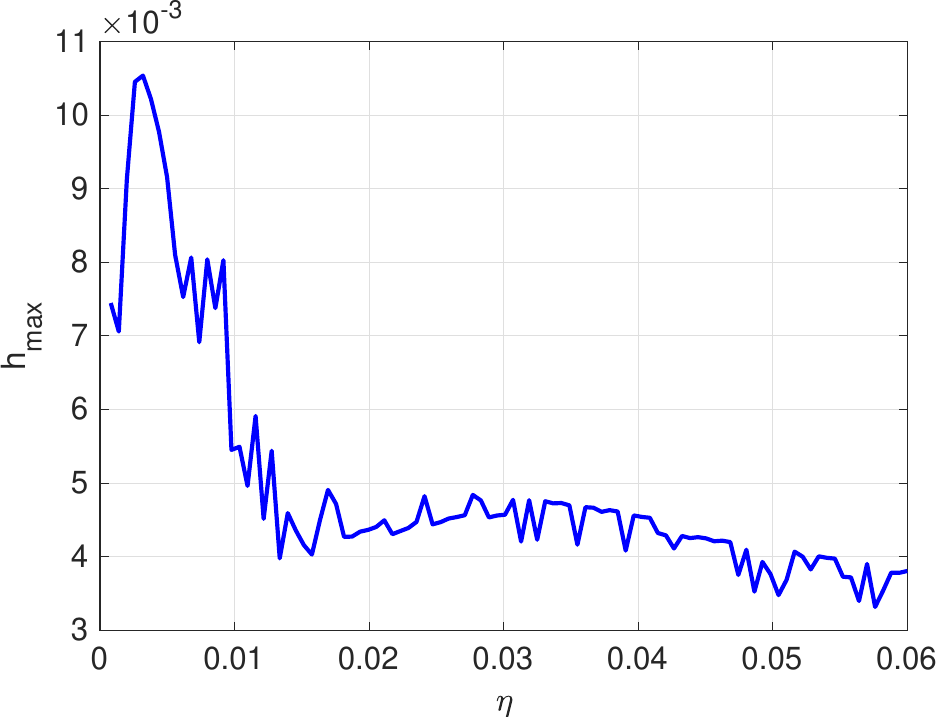}}
    \caption{Study as a function of the dynamic viscosity  $\eta$ (a) of $F_{x,max}$ (b) of $h_{max}$.}
\label{SwapEta}
\end{figure}

\begin{figure}[H]
    \centering
    \subcaptionbox{}
    {\includegraphics[scale=0.4]{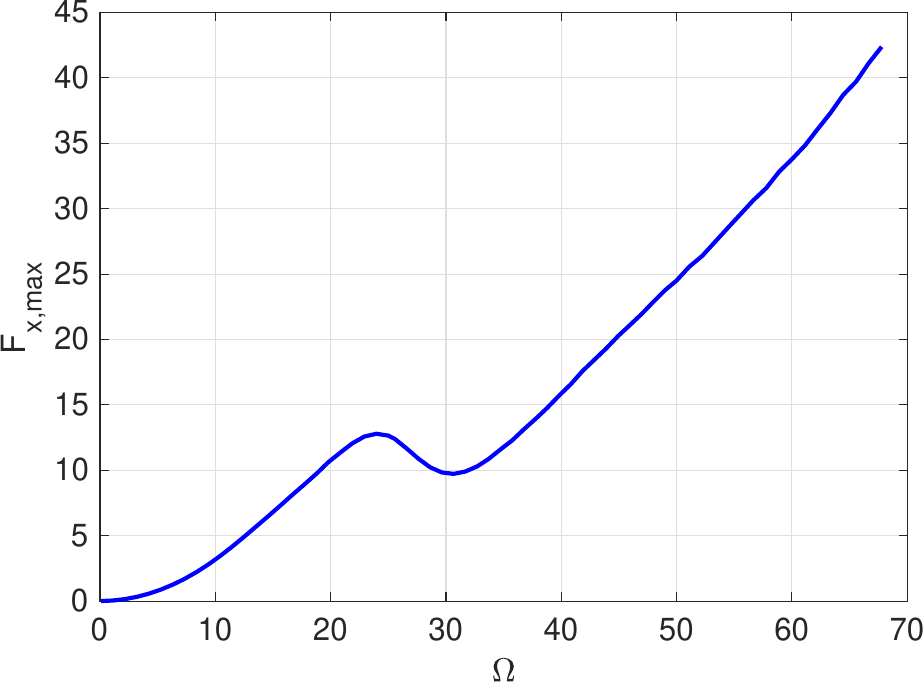}}
    \hfill
    \subcaptionbox{}
    {\includegraphics[scale=0.4]{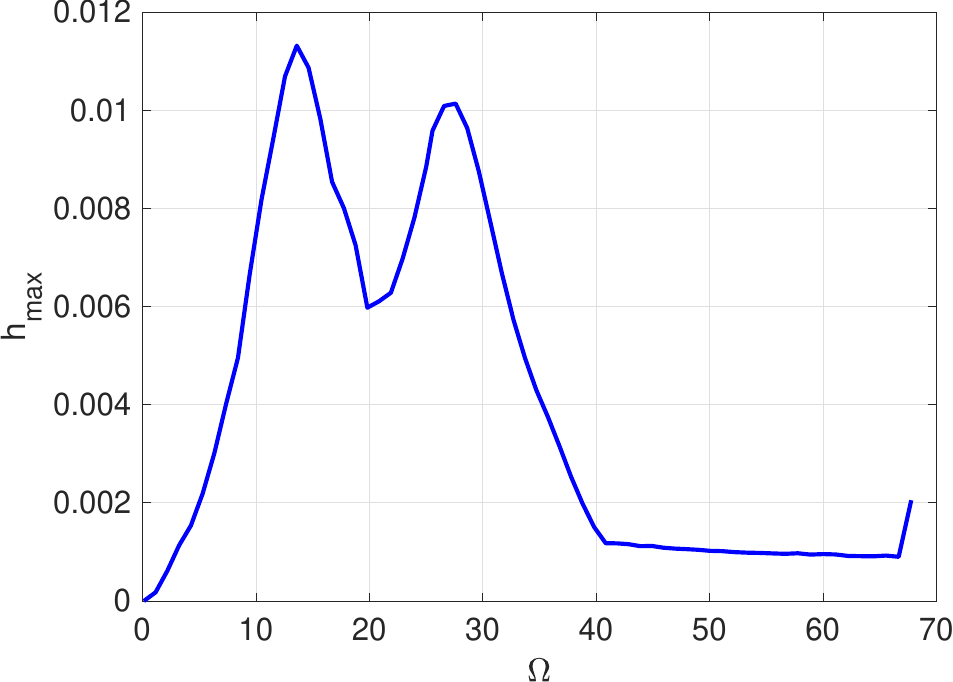}}
    \caption{Study as a function of the pulsation $\Omega$ (a) of $F_{x,max}$ (b) of $h_{max}$.}
\label{SwapOmega}
\end{figure}

\begin{figure}[H]
    \centering
    \subcaptionbox{}
    {\includegraphics[scale=0.4]{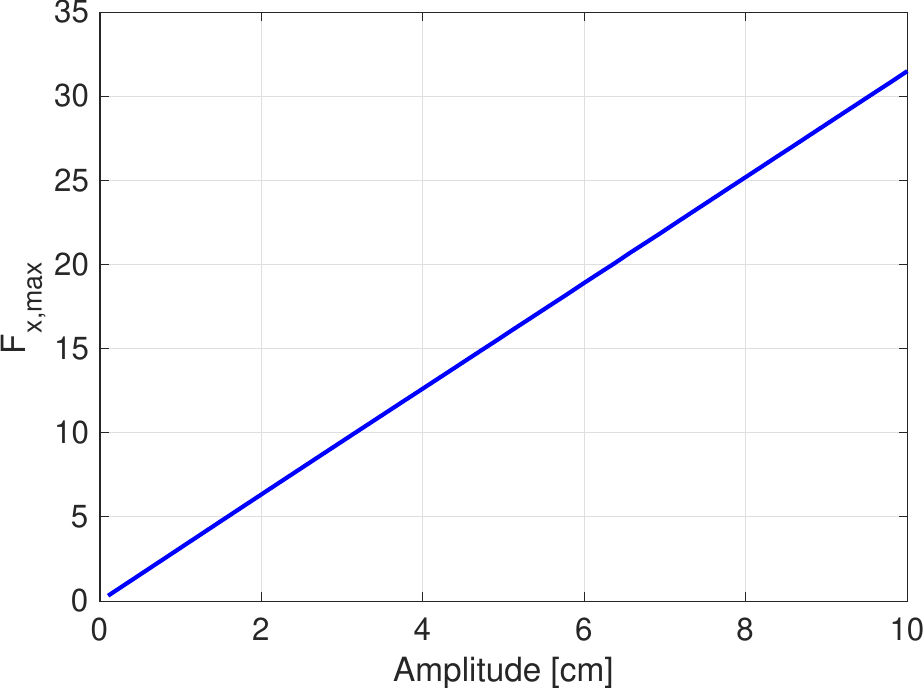}}
    \hfill
    \subcaptionbox{}
    {\includegraphics[scale=0.4]{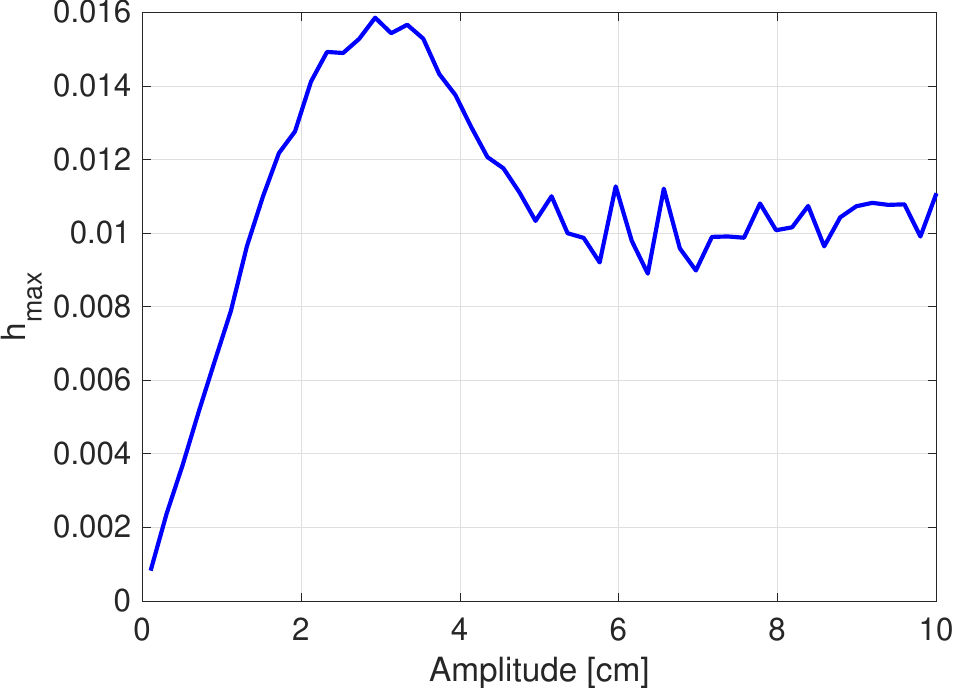}}
    \caption{Study as a function of the excitation amplitude $X_0$ for $\Omega=10\ \rm{rad}\cdot s^{-1}$ (a) of $F_{x,max}$ (b) of $h_{max}$.}
\label{SwapAmlp}
\end{figure}

\chapter{\ Characteristics of liquids}
\label{2_0_0}
The characteristics of the different liquids used in the experiments at $T = 20^{\rm{o}}$C are given in the Table \ref{cara_liq}.
\begin{table}[H]
\centering
\begin{tabular}{|c|c|c|}
\hline
 & Water & Sunflower oil\\
\hline
$\rho$ [kg.m$^{3}$]& $998.30$ \cite{Schmidt1969}& $916.9$ \cite{Bernat2012}\\
\hline
$\nu$ [m$^{2}$.s$^{-1}$] & $1.007 \cdot 10^{-6}$ \cite{Hogge}& $73.45 \cdot 10^{-6}$ \cite{Bernat2012}\\
\hline
$\gamma$ [N.m$^{-1}$] & $72.8 \cdot 10^{-3}$ \cite{Lange} & $33.75 \cdot 10^{-3}$  \cite{NIIR2002}\\
\hline
\end{tabular}
\caption{Density, kinematic viscosity and surface tension of different liquids for a temperature $T = 20^{\rm{o}}$C.}
\label{cara_liq}
\end{table}

\chapter{\ Characteristics of the ERT Bella-Lui rocket}
\label{3_0_0}
\section{Fuel}
The tank is filled with liquid nitrous oxide at a pressure of $P = 69$ bar. At this pressure, it has the following characteristics \cite{Jamieson}:
\begin{itemize}
    \item its temperature is $T = 305 $ K,
    \item its density is $\rho = 660.5$ kg.m$^{-3}$,
    \item its surface tension is $\gamma = 0.5$ mN.m$^{-1}$.
\end{itemize}

\section{Flight characteristics}
\begin{figure}[H]
    \centering
    \subcaptionbox{}
    {\includegraphics[scale=1]{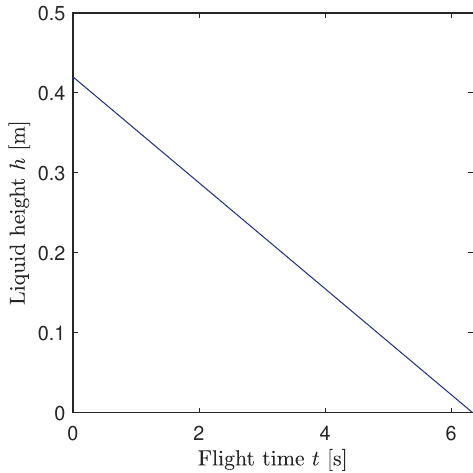}}
    \hfill
    \subcaptionbox{}
    {\includegraphics[scale=1]{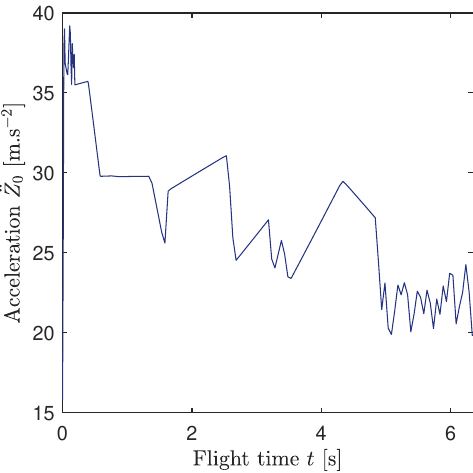}}
    \caption{Evolution (a) of the propellant height $h$ and (b) the vertical acceleration of the rocket $\Ddot{Z}_0$ as a function of flight time.}
    \label{a_p}
\end{figure}

The evolution of the propellant height $h$ and the vertical acceleration of the rocket  $\Ddot{Z}_0$  during the acceleration phase are given in Figures \ref{a_p}(a) respectively \ref{a_p}(b).

\section{Plan of the tank}
\label{plan}
\begin{figure}[H]
    \centering
    \includegraphics[scale=0.6]{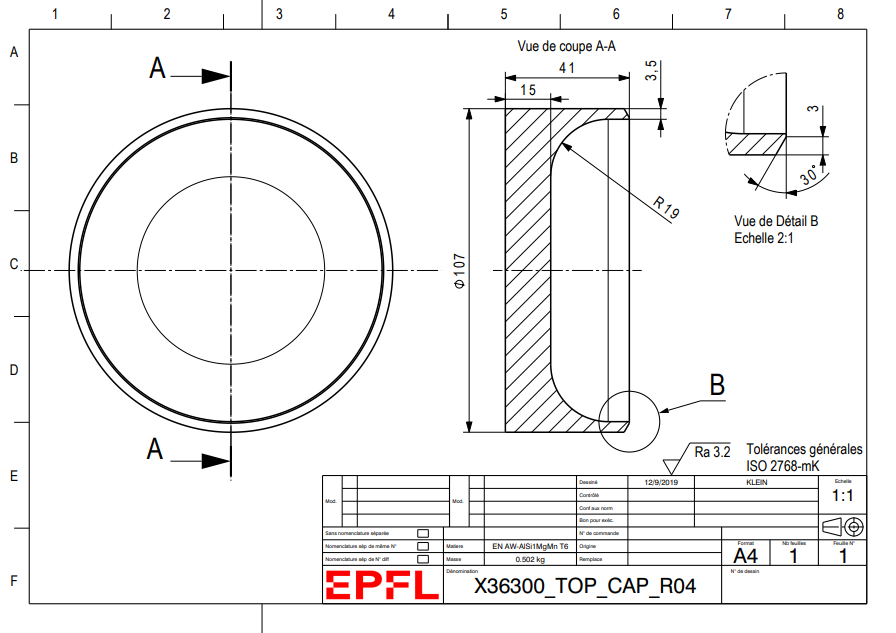}
    \caption{Top view of the Bella-Lui tank}
    \label{plan1}
\end{figure}

\begin{figure}[H]
    \centering
    \includegraphics[scale=0.6]{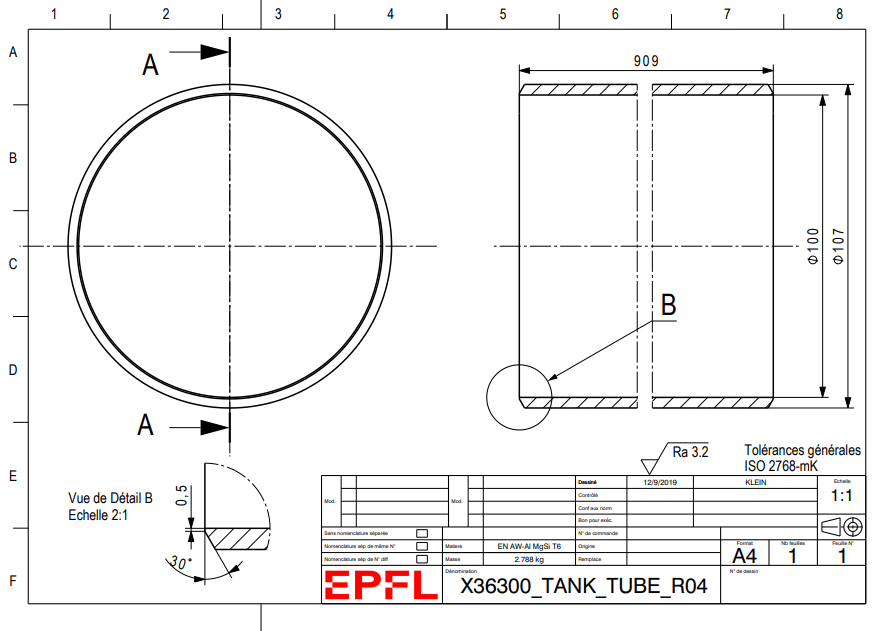}
    \caption{Side plan of the Bella-Lui tank}
    \label{plan2}
\end{figure}

\begin{figure}[H]
    \centering
    \includegraphics[scale=0.6]{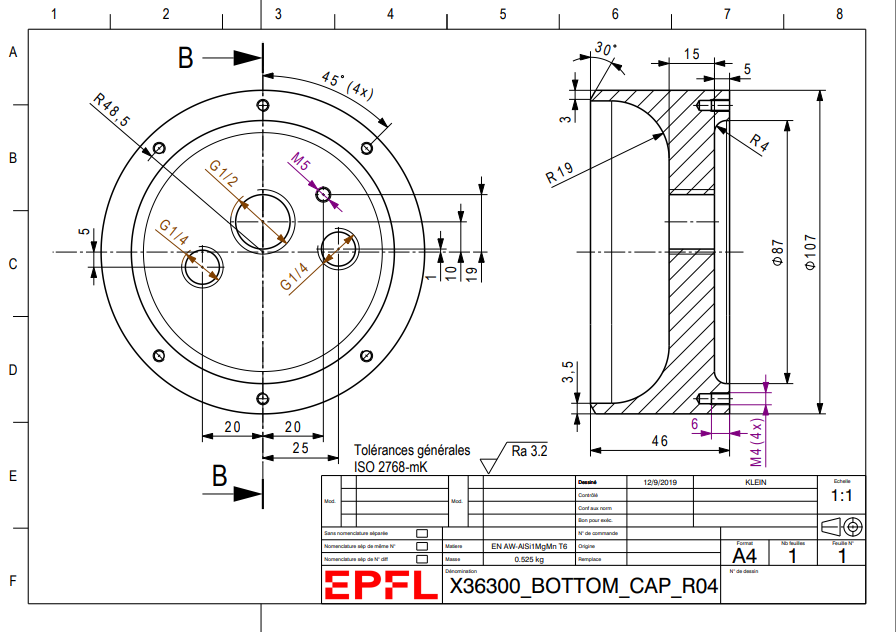}
    \caption{Plan of the bottom of the Bella-Lui tank}
    \label{plan3}
\end{figure}

The plan of the Bella-Lui rocket tank is shown in Figures \ref{plan1}, \ref{plan2}, \ref{plan3}. The tank has the following characteristics :
\begin{itemize}
    \item its radius is $R = 0.057 $ m,
    \item its height is $H = 0.909$ m.
\end{itemize}

\printbibliography

\end{document}